\documentclass[12pt]{article}

\usepackage{PRIMEarxiv}
\usepackage[utf8]{inputenc} % allow utf-8 input
\usepackage[T1]{fontenc}    % use 8-bit T1 fonts
\usepackage{hyperref}       % hyperlinks
\usepackage{url}            % simple URL typesetting
\usepackage{booktabs}       % professional-quality tables
\usepackage{amsfonts}       % blackboard math symbols
\usepackage{nicefrac}       % compact symbols for 1/2, etc.
\usepackage{microtype}      % microtypography
\usepackage{fancyhdr}       % header
\usepackage{graphicx}       % graphics
\usepackage{setspace}
\usepackage{hyperref}
\hypersetup{hypertexnames=false} 
\hypersetup{hidelinks=true} 
\usepackage{amsthm} % for proof
\usepackage[toc]{appendix}
\usepackage[cmex10]{amsmath}
\usepackage{amsfonts}
\usepackage{tabularx}
\usepackage{multirow}
\usepackage{hhline}
\usepackage{fancyhdr}
\usepackage[skip=1\baselineskip]{caption}
\usepackage[authoryear,sort]{natbib}
\usepackage[normalem]{ulem}
\usepackage{algpseudocode}
\usepackage{algorithm}
\usepackage{amsmath,amssymb,mathtools}
\usepackage{bbm}
\usepackage{color}

\newcommand{\Bx}{\mathbf{x}}

\newcommand{\BX}{\mathbf{X}}
\newcommand{\Bz}{\mathbf{z}}

\DeclarePairedDelimiter\abs{\lvert}{\rvert}%
\makeatletter
\let\oldabs\abs
\def\abs{\@ifstar{\oldabs}{\oldabs*}}

\title{Analysing symbolic data by pseudo-marginal methods
}

\author{Yu Yang\thanks{School of Economics, UNSW Sydney, Australia}\:\,\thanks{UNSW Data Science Hub, UNSW Sydney, Australia} \and Matias Quiroz\thanks{School of Mathematical and Physical Sciences, University of Technology Sydney, Australia}\:\,\thanks{Human Technology Institute, University of Technology Sydney} \and Boris Beranger\thanks{School of Mathematics \& Statistics, UNSW Sydney, Australia}\:\,\footnotemark[2] \and Robert Kohn\footnotemark[1]\:\,\footnotemark[2]\:\,\thanks{Data Analytics Center for Resources and Environments (DARE)} \and Scott A.~Sisson\footnotemark[5]\:\,\footnotemark[2] }

\begin{document}

\maketitle

\begin{abstract}
Symbolic data analysis (SDA) aggregates large individual-level datasets into 
a small number of distributional summaries, such as random rectangles or random histograms. The inference is carried out using these summaries in place of the original dataset, resulting in computational gains at the loss of some information. In likelihood-based SDA, the likelihood function is characterised by an integral with a large exponent, which limits the method's utility as for typical models the integral is unavailable in closed form. In addition, the likelihood function is known to produce biased parameter estimates in some circumstances. Our article develops a Bayesian framework for SDA methods in these settings that resolves the issues resulting from integral intractability and biased parameter estimation using pseudo-marginal Markov chain Monte Carlo methods. We develop an exact but computationally expensive method based on path sampling and the Poisson estimator, and a much faster, but approximate, method based on a Taylor expansion. Through simulation and real-data examples we demonstrate the performance of the developed methods, showing large reductions in computation time compared to the full-data analysis, with only a small loss of information.

\end{abstract}

% keywords can be removed
\keywords{Intractable likelihood, pseudo-marginal MCMC, Symbolic data analysis}

\section{Introduction}
Large datasets are one of many contemporary challenges faced by statistical modelling. Recently a range of innovative approaches were developed to reduce computational overheads for modelling with large datasets while minimising or eliminating loss of statistical efficiency. These include 
divide-and-conquer approaches \citep{scott+bbcgm16,nemeth2018merging, wang2023divide}, where data are split across multiple machines, analysed separately, and then the results combined;
core sets \citep{huggins+cb16,campbell2019automated, campbell2019sparse}, where a small weighted subset of the data is identified that contains the core information in the dataset;
data subsampling \citep{bardenet2017markov, quiroz+kvt19, salomone2020spectral}, which iteratively estimates the likelihood based on subsets of the data as part of a Bayesian analysis;
and symbolic data analysis \citep{billard+d03,beranger2018new, whitaker2021logistic}, where subsets of the data are summarised into distributional forms for subsequent analysis. Here we focus on symbolic data analysis (SDA). 

SDA partitions a large dataset into a number of smaller datasets, usually based on some natural grouping within the data, such as conditioning on a relevant explanatory variable \citep{billard+d03}. Each smaller dataset is then reduced to a distributional summary, such as a random interval, random rectangle, or random histogram, describing the approximate distribution of the data \citep{billard2006symbolic}. One direction in SDA takes the distributional summaries themselves -- termed {\em symbols} -- as the objects of interest, models them directly, and produces inference at the symbol level \citep{billard2002symbolic,brito2014symbolic,lin2017estimating}. Another direction, and the one followed here, uses the distributional summaries merely as a convenient way to represent a large complex dataset by lower dimensional quantities. The aim is to fit models that describe the data underlying the original large dataset  -- the {\em micro-data} -- while only observing the distributional summaries \citep{zhang2020constructing,beranger2018new}. The idea is that the computational efficiency gained by working with summary representations of the data may be more beneficial than the loss of statistical efficiency when moving from the micro-data to the summaries.

For $n$ micro-data observations drawn independently from some density, 
in the absence of low dimensional sufficient statistics the standard likelihood function is a product of $n$ terms. 
If these data are summarised as counts within a number of bins, $B\ll n$, then the resulting symbolic log-likelihood function only has $B$ terms \citep{beranger2018new}, with 
each term of the form $n_b \log p_b(\theta)$, where  $p_b(\theta)$ is the probability that a single micro-data point falls in the $b$-th bin, and $n_b$ is the total number of micro-data in bin $b$.
This approach greatly reduces computation (because the number of terms in the likelihood is greatly reduced), and improves data privacy
(in that those 
micro-data in any bin
are indistinguishable from other micro-data in the same bin), in exchange for some loss of statistical efficiency or accuracy (in that the precise location of each micro-data point
is now unknown). 
In the logistic regression context, \cite{whitaker2021logistic} demonstrates that using 
the symbolic likelihood can achieve classification rates compared to the standard full data multinomial analysis and against state-of-the-art subsampling algorithms for logistic regression, but at a substantially lower computational cost.
The symbolic likelihood approach 
is extended to higher-dimensional analyses via composite likelihoods by \cite{whitaker2020composite},
and is applied in 
analyses of credit risk \citep{zhang2020constructing},
computer network traffic modelling \citep{rahman2020likelihood},
crop classification from satellite data \citep{whitaker2021logistic},
and max-stable models for spatial extremes \citep{whitaker2020composite}.

However, despite its flexibility, the approach proposed by \cite{beranger2018new} 
has a number of limitations. Primarily, modelling 
is restricted to density functions with $p_b(\theta)$ available in closed form, which is a strong practical limitation on the range of possible models available (\citealp{beranger2018new} limit their analyses to models based on normal or skew-normal distributions).
This further places limits on the dimension of the micro-data 
%$X$ 
for which the integral is possible \citep[although see][who mitigate dimension issues via composite likelihoods]{whitaker2020composite}.
In addition, in the particular case when representing the micro-data via a single bin
%$\mathcal{B}=\{B_1,B_1^c\}$ 
defined by observed marginal minimum and maximum data values, the symbolic likelihood 
%\eqref{eqn:simpleSymbolicLikelihood} 
produces biased estimates of dependence parameters of the micro-data model $g_X$ (e.g.~correlation parameters in a multivariate normal model) as the number of data points ($n_b$) in the bin becomes large \citep{beranger2018new}.

Our paper makes three important contributions to SDA-based inference. 
First, we extend the class of models implementable via the symbolic likelihood 
to include those where the integral underlying $p_b(\theta)$
is unavailable in closed form, but where an unbiased estimate $\widehat{p}_b(\theta)$ of the integral is available, e.g., via Monte Carlo integration. To achieve this we develop an exact algorithm for Bayesian inference based on signed \citep{lyne2015russian} pseudo-marginal Metropolis-Hastings \citep{andrieu2009pseudo} and correlated pseudo-marginal MCMC via blocking \citep{tran2016block} for the symbolic likelihood. By exact, we mean that posterior expectations with respect to the symbolic data posterior can be estimated to an arbitrarily high precision, provided the MCMC run is sufficiently long. This differs from the usual sense of ``exactness'' in MCMC, which refers to the posterior samples being (asymptotically) distributed according to the symbolic data posterior. Throughout the paper, whenever we refer to our method as exact, it is in this former sense.
We also propose a biased version of these algorithms that achieves faster posterior simulation with low bias. Each of these algorithmic contributions can be more generally applied to models outside of the SDA context, which contain likelihood terms that include intractable quantities with large exponents (i.e., of the form $p(\theta)^n$).
Second, with a minor modification to the symbolic likelihood, we are able to remove the bias in the estimation of dependence parameters for marginal min-max defined single-bin symbols, as observed by \cite{beranger2018new}.
Finally, we demonstrate how to increase the accuracy of inference in the case of single-bin random rectangles 
by combining the symbolic likelihood of observations within the bin with standard likelihood contributions for those observations outside of it, for a moderate computational cost.

This paper is organised as follows. Section~\ref{sec: Methodology} introduces the symbolic likelihood-based framework of \cite{beranger2018new} and our modifications to symbol construction that reduce dependence parameter estimation bias, and increase inferential accuracy for a moderate computational effort. Section \ref{sec:PMCMC} focuses on inference with the modified symbolic likelihood, via exact path sampling, approximate sampling via Taylor expansion with bias-correction, and signed block pseudo-marginal Metropolis-Hastings (PMMH) sampling. Simulation studies in 
Section~\ref{sec: Simulation} demonstrate unbiasedness comparisons of dependence parameter estimates with  \cite{beranger2018new}, a performance comparison of our proposed exact and approximate inferential methods, and a comparison between the approximate inferential method applied to a factor model and the full data result. 
Section~\ref{sec: Empirical study} demonstrates the method on a real dataset using a Bayesian linear regression, and we conclude with a discussion.

\section{Symbolic data analysis} 
\label{sec: Methodology}
\cite{beranger2018new} develop a likelihood-based approach for SDA which incorporates the  process of constructing the symbolic data summaries from the micro-data. This section first explores the construction of this likelihood, its strengths and weaknesses, and then introduces our modifications that both remove dependence parameter estimation bias (observable in some circumstances) and increase inferential accuracy at a moderate computational cost.

\subsection{Problems with SDA likelihoods}
Write the unobserved micro-data $\BX =\{X_1,\ldots,X_n\}$ with $X_i\in\mathcal{X}\subseteq\mathbb{R}^d$, as generated by a density $g_\BX(\BX; \theta)$ with parameter vector $\theta\in\Theta$. When the individual $X_i$ are independent draws, then $g_\BX(\BX; \theta)=\prod_{i=1}^ng_X(X_i; \theta)$. The observed values $\Bx$ of the micro-data $\BX$ are aggregated into a symbol $s=S(\Bx)$. \cite{beranger2018new} discuss the process of how the symbol $s$ is generated from $\Bx$ via the conditional distribution form $f_{S|\BX = \Bx} (s| \Bx)$.
In general, the symbolic likelihood has the form
\begin{equation}\label{eq: symb_lik}
    L(s;\theta) \propto \int_{\mathcal{X}} f_{S|\BX = \Bz} (s|\Bz) g_{\BX} (\Bz;\theta) d\Bz.
\end{equation}
Equation \eqref{eq: symb_lik} states that the symbolic likelihood $L(s;\theta)$ is the marginal distribution of $s$ obtained by integrating over all possible latent (unobserved)  micro-datasets $\Bz$, generated from density $g_{\Bx} (\Bz;\theta)$, that could have generated the observed symbol $s$.

Suppose that in particular we wish to represent the micro-data via the summary $S=S(\BX)=(S_1,\ldots,S_B)$, where $S_b$ is the number of micro-data observations that fall into bin $B_b$, for a set of bins $\mathcal{B}=\{B_1,\ldots,B_B\}$ that partition $\mathcal{X}$. This summary describes a
range of definitions of random intervals, rectangles and histograms. In this case, \cite{beranger2018new} show that the symbolic likelihood function reduces to the form
\begin{equation}
\label{eqn:simpleSymbolicLikelihood}
    L(S; \theta) \propto \prod_{b=1}^B p_b(\theta)^{S_b} \times \ell(\theta),
\end{equation}
where $p_b(\theta)=\int_{B_b}g_X(z; \theta)dz$ is the probability of one $X_i$ falling in bin $B_b$ under the model $g_X(X; \theta)$, and $\ell(\theta)$ is a term that varies according to the precise summary construction.
That is, the symbolic likelihood provides a way to estimate the parameters $\theta$ of the micro-data model $g_X$ based only on the summary $S$.
Particular cases include random histograms where the bins ${\mathcal B}$ are fixed and $S$ is random (for which $\ell(\theta)\propto1$); random histograms where the counts $S$ are fixed and the bins $\mathcal{B}$ are random, e.g.\ when constructing the bins via marginal order statistics (for which $\ell(\theta)=\prod_b g_X(X_{(b)}|\theta)$ for univariate $X$, where $\{X_{(b)}\}$ is the set of order statistics separating the bins); and random rectangles, where $\mathcal{B}=\{B_1, B_1^c\}$ where $B_1$ is a single bin (with $B_1^c$ its complement), perhaps defined by marginal quantiles, that contains a known proportion of the data, and which provides a simple representation of data location and scale.

\begin{figure}[ht!]
    \centering
    \includegraphics[width=1.0\linewidth]{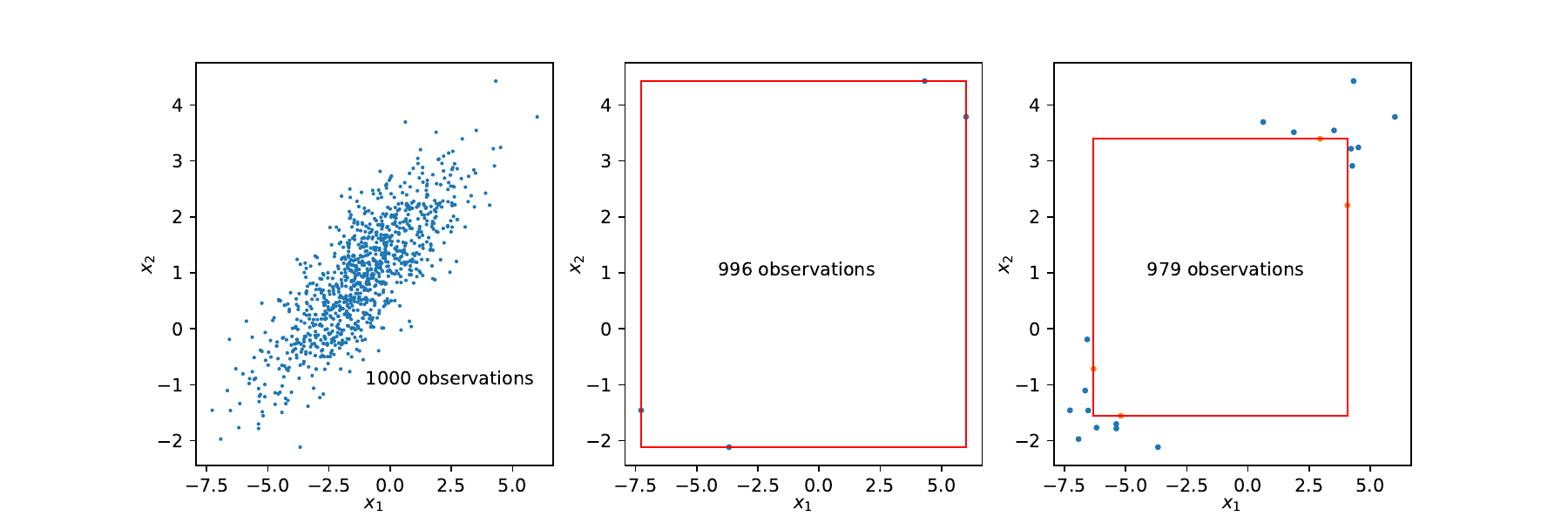}
    \caption{\small Random-rectangle $B_1$ symbol construction. Left panel: Original dataset of $n=1{,}000$ independent observations, $\Bx$, generated from a bivariate normal distribution with $\boldsymbol{\mu}=(-1,1)', \sigma_1=2, \sigma_2=1,\rho_{12} = 0.8 $. Middle panel: Min-max random rectangle with $n_b=4$ points ($\Bx_b$) on the boundary, discarding the locations of the $n-n_b=996$ points within the rectangle.  Right panel: Quantile-based $B_1$ with $q = 0.005$, with $n_b=4$ points on the boundary, $n_e=17$ external points and the remaining $n-n_b-n_e=979$ points inside $B_1$. }
    \label{fig1:symb_constrution}
\end{figure}

We focus on the case where the micro-data are summarised by a single random rectangle $B_1$ defined as the minimal bounding box that contains all of the micro-data (the so-called {\em min-max hyper-rectangle}, defined by the smallest and largest data values in each dimension; see 
Figure~\ref{fig1:symb_constrution}, left and centre panels). This is the most common distributional representation of micro-data within the SDA literature. 
However, most of the results presented here extend to the case of multiple bins (i.e.~histograms).

To be able to identify the dependence parameters of a multivariate normal micro-data model $g_X$ from the min-max hyper-rectangle in addition to location and scale parameters, \cite{beranger2018new} define the random rectangle to comprise the minimum bounding box, the number and indicative (but not precise) position of the micro-data $\Bx_b$ that sit on the boundary of the box (e.g., ``two points, one in the bottom left corner, one in the top right''), and the total number of points ($n$) in $\Bx$.
In this setting, and assuming $g_\BX(\BX; \theta)=\prod_{i=1}^ng_X(X_i; \theta)$ for expositional clarity, the resulting symbolic likelihood function becomes
\begin{equation}\label{eq: boris_symb_lik}
    L_{B_1}(s;\theta) \propto \bigg[ \int_{B_1} g_{X}(z;\theta) dz \bigg]^{n -n_b} \times \mathcal{L}_{b}(\Bx_b; \theta), 
\end{equation}
where $n$ and $n_b$, respectively, denote the total number of observations and the number of observations $\Bx_b$ defining the bounding box. $\mathcal{L}_{b}$ is a likelihood term which takes into account the number and (imprecise) location of the micro-data points $\Bx_b$ used to construct the random hyper-rectangle. The general expression for this term is complex and case-dependent; see the full expression in \citet[Section 2.3.1]{beranger2018new}.

The likelihood in \eqref{eq: boris_symb_lik} can be used to estimate dependence information from the marginal min-max bounding box due to information regarding the number and indicative location of the points $\Bx_b$ used in the box construction in the term $\mathcal{L}_{b}(\Bx_b; \theta)$. E.g., bivariate random rectangles constructed from only two points (bottom-left/top-right or top-left/bottom-right) indicate strong dependence (positive or negative) in the underlying micro-data; rectangles constructed from four data points $\Bx_b$ likely indicate weak or no dependence. (The first term in $L_{B_1}(s;\theta)$ only contains marginal information.)

This symbolic likelihood is computationally efficient compared to the full micro-data likelihood (i.e., the first term accounts for $n-n_b$ micro-data points), and  \cite{beranger2018new} demonstrate that it can estimate correlation parameters $\rho$ when $g_X(X;\theta)=\phi(\mu,\Sigma)$ is specified as a bivariate normal distribution.

However, \cite{beranger2018new} note that $\rho$ is increasingly underestimated as $n$ increases. This occurs as the information in $\Bx_b$ utilised by $\mathcal{L}_{b}(\Bx_b; \theta)$ is imprecise: it does not use the exact location of the bounding points $\Bx_b$. Hence, as $n$ increases, even for extremely strong correlation within $X$, the bounding box is eventually created from four distinct data points, rather than (say) two, which, according to $\mathcal{L}_{b}(\Bx_b; \theta)$, indicates weak or no dependence. If $\mathcal{L}_{b}(\Bx_b; \theta)$ instead makes use of the exact locations of $\Bx_b$ (e.g., noting two points are close together in the far top-right and two close in the far bottom-left, indicating strong dependence), this underestimation is unlikely to occur.

Additionally, the likelihood in \eqref{eq: boris_symb_lik} uses a minimum number of data points $n_b$ to estimate any dependence parameter ($n_b=2,3$ or $4$ for bivariate $X$), which is likely to be extremely inefficient. Using a larger number of micro-data points is likely to improve the estimates, but without losing the efficiency of combining the contribution of large amounts of micro-data into a single likelihood term.

\subsection{Fixing the bias and improving efficiency}

Hence we propose to address both of these issues by slightly reworking the symbolic likelihood \eqref{eq: boris_symb_lik} by 
first redefining the single random rectangle $B_1$ as the
minimal bounding box containing all data in $\Bx$ that are in the central $(1-2q)\%$ of data in all univariate margins, $q\in[0,0.5]$. Figure \ref{fig1:symb_constrution} (right) illustrates this with $q=0.005$, where $B_1$ is the random rectangle constructed from the marginal $nq$ and $n(1-q)$ order statistics. The micro-data can then be partitioned as $\Bx=(\Bx_r,\Bx_b,\Bx_e)$, respectively, containing the $n_r=n-n_b-n_e$, $n_b$ and $n_e$ micro-data points which are in the random interval $B_1$, those which lie on the bounding box, and those which are external to the bounding box. When $q=0$ we recover the original min-max random rectangle ($n_e=0$), and as $q$ increases we allow more of the non-central micro-data to reside outside of $B_1$.

With this partition, we rewrite \eqref{eq: boris_symb_lik} as
\begin{equation}\label{eq2: symb_lik}
    L_{B_1}(s;\theta) \propto \bigg[ \int_{B_1} g_{X}(z;\theta) dz \bigg]^{n -n_b-n_e} 
    \times L(\Bx_b; \theta)
    \times L(\Bx_e; \theta),
\end{equation}
where $L(\BX;\theta)\propto\prod_{i=1}^ng_X(X_i; \theta)$ is the micro-data likelihood. With this construction, the large computational efficiencies are still present as, for small $q$, the first term still contains the vast proportion of the micro-data. However, by replacing the complicated $\mathcal{L}_{b}(\Bx_b; \theta)$ term in \eqref{eq: boris_symb_lik} with the simplified $L(\Bx_b; \theta)$, the full information about the bounding points $\Bx_b$ is included in the likelihood. Further, inclusion of the standard likelihood terms for data outside of the bin, $L(\Bx_e;\theta)$, allows the dependence parameters to be estimated more efficiently. Section \ref{sec: simulation_study1} demonstrates by replicating the simulation study from \cite{beranger2018new} that even with $q=0$ (with $\Bx_e=\emptyset$), so that the bivariate random rectangle is only constructed from $n_b=2, 3,$ or $4$ points, the correlation parameter $\rho$ of the bivariate normal distribution is estimated unbiasedly. As $q$ increases, more micro-data points make their standard contribution to \eqref{eq2: symb_lik}, increasing the micro-data information in the likelihood. In the extreme case of $q=0.5$, all micro-data is included, and \eqref{eq2: symb_lik} reduces to the full micro-data likelihood $L(\Bx;\theta)$. The parameter $q$ thereby represents a balance between computational efficiency (low $q$) and likelihood precision (high $q$). Ideally we are looking for the greatest computational efficiency for acceptable precision. Section \ref{sec: Empirical study} demonstrates this trade-off empirically.

Existing analyses and applications of the likelihood \eqref{eq: boris_symb_lik} are restricted to cases having a closed form integral $\int_{B_1} g_{X}(z;\theta) dz$, or where the dimensionality of $z$ is sufficiently low that a fast and accurate numerical approximation is available \citep{zhang2020constructing,rahman2020likelihood,whitaker2020composite,whitaker2021logistic}. In other cases, an unbiased Monte Carlo estimate is always available via $\int_{B_1} g_{X}(z;\theta) dz\approx\frac{1}{N}\sum_ig(z_i;\theta)$ where the $z_i\in B_1$ are sampled uniformly over the bin. However, raising this estimate to a fixed positive power in \eqref{eq: boris_symb_lik} or \eqref{eq2: symb_lik} produces a biased estimate of the likelihood (and log-likelihood) function. In the next section, we develop exact and (faster) approximate samplers that permit unbiased and approximately unbiased posterior simulation in this setting.

\section{Likelihood estimators}

\label{sec:PMCMC}

For $X$ with more than 2 dimensions ($d >2$), the analytical evaluation of the integral in \eqref{eq2: symb_lik} is often unavailable even for well-known distributions. The likelihood $L_{B_1}(\theta)$ is of the form $p(\theta)^n$ (omitting the micro-data terms) with large $n\in\mathbb{Z}^+$, where $p(\theta)$ is a probability that can be estimated unbiasedly by $\widehat{p}(\theta)$ obtained e.g.\ via Monte Carlo integration.  However, using this estimate directly within the likelihood as $\widehat{L}_{B_1}(\theta) = \widehat{p}(\theta)^{n}$  or $\widehat{\log} L_{B_1}  = n \log \widehat{p}(\theta)$  results in a biased estimate of the likelihood or log-likelihood function, which is problematic when used within a pseudo-marginal Metropolis-Hastings (PMMH) sampler, which typically requires an unbiased, positive estimate \citep{andrieu2009pseudo}. In practice the likelihood function may comprise many such terms $\widehat{L}(\theta) = \prod_{b=1}^B\widehat{p}_b(\theta)^{n_b}$, when the data are aggregated over several random rectangles or bins in a random histogram. Inspired by the work of \cite{gelman1998simulating} and \cite{papaspiliopoulos2011monte}, Section \ref{sec: exact method} proposes an exact path-sampling method for unbiased estimation of the symbolic likelihood in \eqref{eq2: symb_lik}. However, this exact method is slow due to the many Monte Carlo simulations involved. Section \ref{sec: approx method} proposes an approximate method to speed up the computation. When the density $g_X$ is a multivariate normal distribution, Section \ref{sec: likelihood est for normal} presents a modification of the minimax exponentially tilted (MET) estimator of \cite{botev2016normal} for efficient estimation of $p(\theta)$.
Sections \ref{sec: Simulation} and \ref{sec: Empirical study} adopt this estimator for approximating $\int_{B_1} g_{X}(z;\theta) dz$ when $g_X$ is Gaussian. Finally, Section \ref{section: pm-mcmc} presents the sampler for our exact method.

\subsection{An exact method: Path sampling using the Poisson estimator}
\label{sec: exact method}

We propose a two-step procedure to construct an unbiased estimator of the symbolic likelihood function in \eqref{eq2: symb_lik}. The first step uses path sampling \citep{gelman1998simulating} to obtain an unbiased estimator of the logarithm of the likelihood. The second step transforms the unbiased estimator of the log of the likelihood to an unbiased estimator of the likelihood by using the Poisson estimator \citep{papaspiliopoulos2011monte}. We now describe the approach in detail.

The logarithm of the symbolic likelihood function (with additive terms not depending on $\theta$ omitted) is 
\begin{equation}
\label{eqn:loglikbit}
        l_{B_1}(s;\theta) \propto (n - n_b - n_e) \log \left[ \int_{B_1} g_{X} (z;\theta) dz \right]+ \log L(\Bx_{b};\theta) + \log  L(\Bx_{e};\theta),
\end{equation}
where the last two terms are standard tractable log-likelihood terms that are ignored in the following discussion.  It is possible to obtain an unbiased estimator for $\int_{B_1} g_{z} (z;\theta) dz $ in the first term via Monte Carlo or other sampling methods. 
  However, unbiasedness is not preserved by the log transformation. Instead, we use the path sampler \citep{gelman1998simulating} to obtain an unbiased estimator of $l_{B_1}(s;\theta)$ as follows. Let $h_t(z; \theta) = g_{X}(z; \theta)^t$, $t \in [0,1]$ and  $$q_t(z;\theta) = \dfrac{h_t(z;\theta)\mathbbm{1}(z\in B_1)}{\int_{B_1} h_t(z;\theta) dz} $$
  be the normalised density of $g_{X}(z; \theta)^t$ truncated to $z\in B_1$. In our paper, we focus on settings where sampling from the truncated multivariate density $q_t(z;\theta)$ is feasible. One such setting arises when $g_X(z;\theta)$ is multivariate normal, since the resulting $q_t$ is then a truncated multivariate normal density that can be sampled using the method in \cite{botev2016normal}.
  
  Following Appendix \ref{app: ps}, the first term in \eqref{eqn:loglikbit} can be expressed as 
\begin{equation}\label{eq: path_sampler}
    (n -n_b - n_e)\log \int_{B_1} g_{X}(z;\theta) dz = (n-n_b - n_e)\bigg(\int_0^1 \mathbb{E}_{q_t(z;\theta)} \bigg[ \frac{d}{dt}\log h_t(z;\theta) \bigg]dt  +\log \int_{B_1} 1 dz\bigg).
\end{equation}

Based on \eqref{eq: path_sampler}, the path sampler offers an elegant way to obtain the desired estimator by integrating over the so-called temperature $t$, which is a one-dimensional quantity. Path sampling reduces the problem of computing a high-dimensional integral to evaluating a single integral over temperature, with the integrand given by expectations under intermediate distributions that can be unbiasedly estimated. This, combined with the key feature of the path sampler of moving the logarithm inside the expectation, yields an unbiased estimator of the log-integral (up to trapezoidal error). The integral 
\begin{equation}\label{eq:integral}
    \int_0^1 \mathbb{E}_{q_t(z;\theta)} \bigg[ \frac{d}{dt}\log h_t(z;\theta) \bigg]dt
\end{equation}
usually does not have an analytical solution, but it is efficiently approximated using numerical integration methods.
Unbiased estimation of \eqref{eq:integral} can be achieved via e.g.~the method in \cite{rischard+jp18}, though for simplicity we use the trapezoidal rule, and note that the unbiasedness of the final path sampler estimate of the true log-likelihood depends on the trapesoidal rule error being vanishingly small.
%We use the trapezoidal rule to this end. 

\begin{figure}[ht!]
    \centering
    \includegraphics[width=0.7\linewidth]{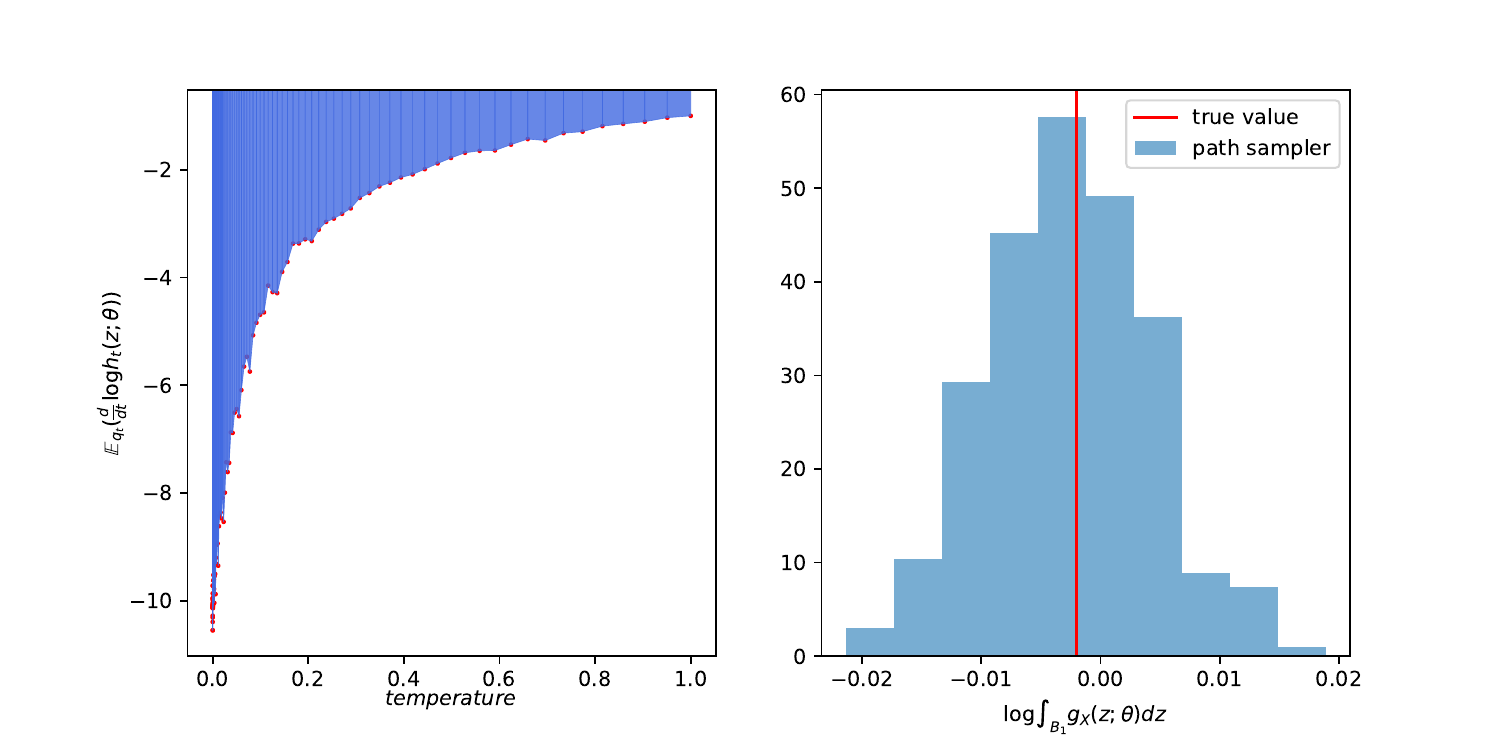}
    \caption{\small Path sampler performance in estimating \eqref{eq:integral}. The model is the same as 
    %the middle plot of 
    Figure  \ref{fig1:symb_constrution}, with the parameters taking their true values. The temperature ladder is set as $t = (i/T)^5, i = 1,\dots T, T = 100$. For each temperature $t$, the expectation is estimated from $M=2{,}000$ $z$'s drawn from $q_t(z;\theta)$. The left panel shows the estimated integrand of \eqref{eq:integral} at each temperature, $t$. The right panel shows a histogram of 500 independent replications of the resulting log-likelihood estimate and the theoretical true value.
    }
    \label{fig2: path sampler}
\end{figure}

To select an appropriate
sequence of temperatures $t \in (0,1]$ for this integral, the so-called temperature ladder, we follow  \cite{friel2008marginal} and specify $t = (i/T)^5, i = 1,\dots,T$. This geometric series fixes the total number of temperatures $T$ and places more points at the lower temperatures, where the value of $\mathbb{E}_{q_t(z;\theta)} \bigg[ \frac{d}{dt}\log h_t(z;\theta) \bigg] $ changes drastically with $t$. Figure \ref{fig2: path sampler} demonstrates the performance of the path sampler using this temperature ladder. The left panel depicts the estimated integrand of \eqref{eq:integral} at each temperature, with the area of the trapezoids in the left panel used for the numerical integration over temperatures. The right panel empirically verifies that the path sampler provides an unbiased estimate (up to trapezoidal rule error) of the true log-likelihood component $\log\int_{B_1}g_X(z;\theta)dz$. In Appendix \ref{app:variance_scaling}, we show heuristically that the variance of the estimator of \eqref{eq:integral} scales as $\mathcal{O}(d)$, with $d=\dim(x)$, under simplifying assumptions. The key assumptions are that $g_X(z;\theta)$ is a multivariate Gaussian with independent components, and that $$q_t(z;\theta) \propto \prod_{i=1}^d N(z_i|\mu_i,\sigma^2_i)^t \mathbbm{1}_{[a_d,b_d]}(z_i),\quad \text{with }  a_d = -c\sqrt{d}, b_d =+c\sqrt{d},  $$ 
where $N(\cdot | \mu, \sigma^2)$ denotes the univariate normal density with mean $\mu$ and variance $\sigma^2$, and $\mathbbm{1}_{A}(x)=1$ is the indicator function of the set $A$. The truncation region (the box) grows with $d$, ensuring that the probability of the tempered Gaussian lying in the box does not collapse to zero as $d$ increases; see Appendix \ref{app:variance_scaling} for details.

Algorithm \ref{algo: path sampler} describes the implementation of the path sampler. 

\begin{algorithm}[ht!]
\caption{The path sampling algorithm} \label{algo: path sampler}
\begin{algorithmic}[1]
\State \textbf{Input}:
\begin{itemize}
    \item[] $B_1$: the known region of the random rectangle;
    \item[] $\theta$: likelihood parameter;
    \item[] $0<t_1<\ldots<t_T=1$: a temperature sequence of length $T$ between 0 and 1;
    \item[] $M$: number of samples to draw at temperature $t_i$, $i\in \{1,\dots,T\}$;
\end{itemize}
\State \textbf{Output}: an unbiased estimator of $\log \int_{B_1} g_{X}(z;\theta)dz$.
\For{$i=1 \to T$}
    \For{$m=1 \to M$}
    \State Sample $z_m$ from $q_{t_i}(z;\theta)$.
    \EndFor
    \State $ \mathcal{T}_{t_i} \gets  
    \frac{1}{M}\sum_{m=1}^M\dfrac{d}{dt_i}\log h_{t_i}(z_m;\theta)$
    \Comment{$\mathcal{T}_{t_i} $: an unbiased estimator of $\mathbb{E}_{q_t(z;\theta)} \big[ \frac{d}{dt_i}\log h_{t_i}(z;\theta)  \big] $.}
\EndFor
\State  Numerically integrate $\mathcal{T}_{t_i}$ from $t_1$ (a number close to 0) to $t_T$ and use \eqref{eq: path_sampler} to obtain the final result.
\end{algorithmic}
\end{algorithm}

Path sampling provides an unbiased estimator (up to trapezoidal rule error) for the logarithm of the likelihood function. The next step is to transform this to the original (likelihood) scale.  Let  $A = (n -n_b - n_e)\log \int_{B_1} g_{X}(z;\theta) dz$, and denote the corresponding path sampling estimator as $\widehat{A}_P$, with $\mathbb{E}[\widehat{A}_P]=A$. Due to the non-linearity of the exponential transformation, $\mathbb{E}(\exp(\widehat{A}_P)) \neq \exp(A)$. The Poisson estimator \citep{papaspiliopoulos2011monte}, denoted as $\widehat{\exp(A_P)}$, ensures that $\mathbb{E}(\widehat{\exp(A_P)}) = \exp(A)$ with 
\begin{equation}\label{eq: poiss_est}
    \widehat{\exp(A_P)} = \exp(a+\lambda) \prod_{h=1}^{\chi}  \dfrac{(\widehat{A}_P^{\,(h)} - a)}{\lambda},
\end{equation}
where $\chi \sim \text{Poisson}(\lambda)$, $\widehat{A}_P^{\,(h)}$, $h = 1,\dots, \chi$, are independent (path sampler) estimates of $A$, and $a$ is an arbitrary real number. It can be shown that, for a fixed $\lambda$, $a = A - \lambda$ minimises the variance of the Poisson estimator (see Appendix \ref{app: pois est}). Note that if $a$ is a lower bound of $\widehat{A}_P$, then the estimator in \eqref{eq: poiss_est} is positive with probability 1. However, it is usually difficult to obtain a tight lower bound, and \cite{quiroz2021block} show that a loose lower bound results in a highly variable estimator. \cite{quiroz2021block} propose to instead use a soft lower bound $a = A_P^* - \lambda$, which they define as a bound that results in a positive estimator with a high probability. For symbolic data problems, the soft lower bound  $A_P^* = (n-n_b -n_e) \gamma^d  d \log (1-2q)$, where all quantities except $\gamma$ have been previously defined, works well in practice. The constant $\gamma^d$, with $\gamma < 1$, is included as we found that $d\log(1-2q)$ usually underestimates $\log \int_{B_1} g_X(z;\theta) dz \leq 0$. We emphasize that the method is exact regardless of the choice of the lower bound.

We use the Poisson estimator within the block pseudo-marginal framework; see Section \ref{section: pm-mcmc} for details. This is because \cite{quiroz2021block} require a large number of products in order to induce a high correlation in the block-pseudo marginal scheme. However, having a large number of products is infeasible in our setting because the path sampler is very computational expensive. Finally, note that when $\chi = 0$, the Poisson estimator in \eqref{eq: poiss_est} reduces to $\exp(a+\lambda)$ , which can be a poor estimator of $\exp(A)$ if the soft lower bound ($a + \lambda = A_P^*$) is not tight (i.e.\ far from $\log \int_{B_1} g_X(z;\theta) dz$). To avoid a high probability of getting $\chi = 0$, we found that $\lambda=3$, which gives $\Pr(\chi = 0) \approx 0.05$, works well in practice.

\subsection{The approximate method: Taylor expansion with bias-correction}

\label{sec: approx method}

Even though the exact method generates an unbiased likelihood estimator, it is computationally costly as there are three nested loops in its implementation: To execute the path sampler (Algorithm \ref{algo: path sampler}), a loop over $T$ temperatures is required. For each temperature, another loop includes $M$ draws for evaluating the expectation. Furthermore, the Poisson estimator requires an average of $\lambda$ replicate unbiased estimators \eqref{eq: poiss_est}.  The computational complexity of $TM\lambda$ is more of a concern in the Bayesian context as the nested loop requires re-evaluation for each parameter proposal in every Monte Carlo iteration. We now develop an approximately bias-corrected estimator, that is much faster than the exact estimator, at the cost of an additional assumption and a narrower scope of applicability.

Similarly to the exact method, the approximate method is also a two-step procedure. In the first step, the logarithm of the likelihood function is approximated by a quadratic Taylor series.  Suppose that $M$ replicate estimators, $\widehat{C}^{(1)},\ldots,\widehat{C}^{(M)}$, are available such that $\mathbb{E}(\widehat{C}^{(m)}) = C = \int_{B_1} g_{X}(z;\theta) dz$, $m = 1,\dots,M$. 

The (stochastic) Taylor expansion of $\log \widehat{C}^{(m)}$ in a neighbourhood of $C$ is
\begin{equation}\label{eq:Taylor_expansion}
    \log \widehat{C}^{(m)} = \sum_{j=0}^2 \frac{\log^{(j)}(C)}{j!} (\widehat{C}^{(m)}- C)^j + o_p((\widehat{C}^{(m)}-C)^2),
\end{equation}
where $\log^{(j)}(C)$ denotes the $j$th derivative of $\log(C)$ with respect to $C$.

Discarding terms of higher order degree than 2
and taking the expectation on both sides of \eqref{eq:Taylor_expansion}, 
\begin{align*}
    \mathbb{E}(\log \widehat{C}^{(m)}) &\approx \log C + \frac{1}{C} \mathbb{E}(\widehat{C}^{(m)}-C)  - \frac{1}{2 C^2} \mathbb{E}((\widehat{C}^{(m)} - C)^2)\\
     &= \log C -  \frac{1}{2 C^2} \mathrm{Var}(\widehat{C}^{(m)}).
\end{align*}
The above equation suggests the following estimator of $\log C$,
\begin{equation}\label{eq: Taylor}
     \widehat{\log C} = \frac{1}{M} \sum_{m=1}^M \log \widehat{C}^{(m)} + \frac{1}{2 \left(M^{-1}\sum_{m=1}^M \widehat{C}^{(m)}\right)^2} \mathrm{Var}(\widehat{C}),
\end{equation}
where the $\widehat{C}^{(m)}$ are replicates of the random variable $\widehat{C}$. Recalling that $A =(n -n_b - n_e)\log \int_{B_1} g_{X}(z;\theta) dz$,
the approximate (Taylor) estimator $A_T$ of $A$ is
\begin{equation} \label{eq: Ahat}
A_T = (n -n_b - n_e)\widehat{\log C}.
\end{equation}
Because $A_T$ is biased for $A$, we cannot use the Poisson estimator with $A_T$ to produce an unbiased estimator of the likelihood as outlined in the previous section.  Instead, we propose the following approach which
%The following approach 
is closely related to the penalty method in \cite{ceperley1999penalty}. Suppose that $\log \widehat{C}^{(m)} $ is approximately normal distributed, i.e.\ $\log \widehat{C}^{(m)} \stackrel{\mathrm{iid}}{\sim}  N(\mu, \sigma^2) $. It follows from the properties of the log-normal distribution that $ \mathbb{E}(\widehat{C}^{(m)}) =\exp(\mu + \sigma^2/2)$. Then, $\widehat{\log C}$ is also likely to be approximately normal as it is a linear combination of $\log \widehat{C}^{(m)}$ plus a constant $\dfrac{1}{2C^2} \mathrm{Var}(\widehat{C})$. Hence, from \eqref{eq: Ahat}, $A_T$ is also (approximately) normally distributed. We may then propose the approximate estimator (the ``approximately bias-corrected estimator''), $\widehat{\exp (A_T)}$ of $\exp(A)$ as:
\begin{align}
    \widehat{\exp(A_T)} = \exp \left(A_T - \frac{1}{2} s(A_T)\right), \label{eq:approximate_likelihood_estimator}
\end{align}
where the sample variance $s(A_T)$ to replaces the unknown $\sigma^2 = \mathrm{Var}(A_T)$, given a relatively large $M$. The Monte Carlo estimates $\widehat{C}^{(m)}$ in \eqref{eq: Taylor} are used to compute the sample variance of $s(A_T)$ based on \eqref{eq: Ahat}. 

Compared to the exact method based on the Poisson estimator, the additional assumption is that the estimator $\widehat{C}^{(m)}$ is log-normal (equivalently, that $\log\left(\widehat{C}^{(m)}\right)$ is normally distributed). Moreover, evaluating $\widehat{C}^{(m)}$ entails unbiasedly estimating a constrained high-dimensional integral, which is challenging to perform efficiently in general. Consequently, the approximate method is restricted to settings where such unbiased estimators are available and computationally efficient, for example, for the multivariate normal \citep{botev2016normal} and Student-t \citep{Botev2022} distributions.

\subsection{The likelihood estimator for a truncated multivariate normal distribution}

\label{sec: likelihood est for normal}

The approximate method in the previous section assumes the availability of unbiased estimators $\widehat{C}^{(m)}, m=1,\dots, M$,  for the integral $C = \int_{B_1} g_X(z;\theta)dz$. Despite taking the average of the independent samples ($\frac{1}{M}\sum_{m=1}^M \widehat{C}^{(m)}$), the estimator of $C$ may still have large variability, especially when the dimension of $X$ (i.e.~$d$) is high since the region (here $B_1$) is restricted. If the underlying distribution $g_X(x;\theta)$ is of a specific form, there may exist a more efficient way of doing the Monte Carlo integration over the bin $B_1$. If we assume that $g_{X}(x;\theta)$ is a $d$-dimensional multivariate normal density function with mean $\mu$ and covariance $\Sigma$ ($\theta = \{\mu,\Sigma\}$), we can use the separation-of-variables (SOV) estimator \citep{genz1992numerical} to evaluate the integral by decomposing the region $B_1$ into $d$ one-dimensional areas. \cite{botev2016normal} extended the SOV estimator to the minimax-exponentially-tilted (MET)  estimator for simulating independent observations from a truncated multivariate normal distribution, as well as computing the cumulative distribution function. The MET estimator has lower variance than the SOV estimator based on simulation studies \citep{botev2016normal}, which is desirable in the current pseudo-marginal MCMC framework. 
Our paper implements the MET estimator for approximating $\int_{B_1} g_{X}(z;\theta) dz$ when $g_X$ is Gaussian. See Appendix~\ref{app: sov} for details.

\subsection{The signed block PMMH algorithm with the Poisson estimator}

\label{section: pm-mcmc}

A popular way of constructing MCMC samplers involving intractable likelihood functions is to use unbiased likelihood estimates in place of the unavailable likelihood function via the pseudo-marginal (PM) method of   \cite{andrieu2009pseudo}. The key criterion for an efficient PM method is that the variances of the logarithm of the likelihood estimates should be sufficiently small, approximately within the interval $[1, 3]$, to achieve an optimal trade-off between sampling efficiency and computational cost \citep{pitt2012some, doucet2015efficient, sherlock2015efficiency}. These guidelines are derived for settings outside of our estimators; however, avoiding a too large variance of the logarithm of the likelihood estimator is crucial for any implementation of PM to avoid the chain getting stuck. This may be particularly hard in the symbolic likelihood setting: for example, one of the contributing terms to the variance of the log of Equation \eqref{eq:approximate_likelihood_estimator} is
$$\mathrm{Var}(A_T) = (n - n_b - n_e)^2 \mathrm{Var}(\widehat{\log C}),
$$
which increases quadratically with the number of observations. Hence, we need to ensure a sufficient number of random samples is used to generate the likelihood estimates, so that $\mathrm{Var}(\widehat{\log C})$ is small enough.

The randomness in the likelihood estimates for $L_{B_1}(s;\theta)$ is determined by the random numbers $u = (u_1,\dots, u_M)$ used to generate the $z_m$ at Step 5 in Algorithm \ref{algo: path sampler} for the path sampler estimator, or the $\widehat{C}^{(m)}$ estimators in \eqref{eq: Taylor} for the approximate method. We denote the estimated likelihood as $\widehat{L}_{B_1}(s;\theta,u)$. Correlating the logarithm of the likelihood estimators at the current and proposed draws increases the sampling efficiency of PM methods by controlling the variability of the likelihood ratio in the Metropolis-Hastings acceptance probability \citep{deligiannidis2018correlated,tran2016block}. We follow the block-pseudo marginal framework in \cite{tran2016block}, 
who group the random numbers $u$ into $U$ blocks, and update the random numbers in a single block jointly with $\theta$, holding the random numbers in the other blocks fixed. The proposed blocked random numbers are denoted by $u'$. The blocking strategy induces a correlation between the logarithm of the likelihood estimates (determined by $u$ and $u'$) in the Metropolis-Hastings acceptance probability. Although we cannot quantify the correlation directly as in \cite{tran2016block}, because our estimators fall outside of their ``product of estimators'' framework. Nevertheless, this does not have implications for the correctness of the results; the main reason for blocking is still the same, i.e.\ to induce a large enough positive correlation to avoid the MCMC chain getting stuck. This is achieved in all our examples and we now outline how to block the random numbers for the exact and approximate methods. See the applications for specific choices of the number of blocks $U$.

For the exact method (Section \ref{sec: exact method}), blocks can be created by grouping at the temperature level, or over the particles drawn at each temperature. For the approximate method (Section \ref{sec: approx method}), where there are $M$ particles involved in generating the likelihood estimate, each particle can form its own block, in which case $U=M$. In either case, because $u$ and $u'$ differ by a few blocks only, the correlation between the logs of the likelihood estimators at the current and the proposed parameter values is likely to be high.

A final issue is that in the exact method, the Poisson estimator in \eqref{eq: poiss_est} sometimes generates a negative estimate for the likelihood function. To cope with this, we follow \cite{lyne2015russian} and instead of using \eqref{eq: poiss_est} we take its absolute value $|\widehat{\exp(A_P)}|$. The final absolute value of the likelihood estimate for one symbol is $|\widehat{L}_{B_1}(s;\theta, u)| \propto |\widehat{\exp(A_P)}| \times L(\Bx_{b};\theta)  \times L(\Bx_{e};\theta) $. As an estimator of $L_{B_1}(s;\theta)$, $|\widehat{L}_{B_1}(s;\theta, u)|$ is no longer unbiased, but simulation-consistent posterior inference is achieved by using the signs of the likelihood estimates when computing posterior expectations \citep{lyne2015russian}. Algorithm \ref{algo: PM-MCMC_SDA} in Appendix \ref{app: alg2} outlines the steps in detail. See e.g.~\cite{lyne2015russian, quiroz2021block} for more details on the so-called signed PMMH algorithm.

\section{Simulations} 
\label{sec: Simulation}
In the following simulation studies, we specify the proposal distribution $q(\theta'|\theta)$ in Algorithm~\ref{algo: PM-MCMC_SDA} as a Gaussian random walk with variance $\sigma\Sigma$. Here $\Sigma$ is constructed using the adaptive random walk strategy in \cite{haario2001adaptive}, and $\sigma$ is adaptively tuned using the Robbins-Monro approach in \cite{garthwaite2016adaptive} so that the sampler achieves a prescribed overall acceptance probability, here $0.234$. See \cite{garthwaite2016adaptive} for further details.
\subsection{Example 1: Estimating the correlation in a bivariate normal distribution}
\label{sec: simulation_study1}

We first compare the ability of our reworked symbolic likelihood construction  $L_{B_1}(s;\theta)$ in \eqref{eq2: symb_lik} to code dependence information, against the original formulation \eqref{eq: boris_symb_lik} of \cite{beranger2018new}, denoted $L_O(s;\theta)$. We replicate the simulation in \citet[Section 3.2]{beranger2018new} in which $m=20,50$ bivariate min-max hyperrectangles are each constructed from $n=5, 10, 50, 100, 1{,}000, 100{,}000$ samples from a bivariate normal distribution with $\mu =(2, 5)^\top$, $\sigma^2_{1}, \sigma^2_{2} = 0.5^2$, and  varying correlations $\rho = 0, 0.3, 0.5, 0.7, 0.9$. Here $n_e=0$ and $n_b=2,3,4$ (randomly), and interest is in estimating the correlation $\rho$ via maximum likelihood. Exact (not estimated) likelihoods are used.

Table \ref{table: bvn} reports the mean and standard deviation of the maximum likelihood estimate $\widehat{\rho}$ under both likelihoods, taken over 100 replicate datasets. 

The results for $L_O$ correspond to those reported in \cite{beranger2018new}. While, for small sample sizes $n$, the estimates are close to the true correlation values, the estimates fall away to zero as the sample size increases. This is most apparent for smaller true correlations; however, the fall-off occurs for all correlations when $n$ becomes large enough. \cite{beranger2018new} argue that when $n$ is large, for any fixed correlation it is increasingly likely that the bivariate min-max  random rectangle is constructed from 4 unique data points. As the number of points on the rectangle boundary, $n_b$, is used by \cite{beranger2018new} to determine the strength of the correlation, (i.e.\ 2 points imply strong correlation; 4 points imply weaker or no correlation), this means that once $n$ is sufficiently large, $L_O$ underestimates the magnitude of $\rho$ for any correlation value. Table \ref{table: bvn} clearly shows this effect, particularly for smaller $\rho$. 

In contrast, for our reworked construction, $L_{B_1}$, the estimates are unbiased and close to the true correlation values in all settings, the estimates have a smaller standard deviation compared to those using $L_O$, and the functional form
%\footnote{what does this mean?} 
of the likelihood term $L(\Bx_b;\theta)$ in \eqref{eq2: symb_lik} -- here simply a product of $n_b$ independent density terms -- is functionally
%\footnote{what does this mean?} 
far simpler than the equivalent term for $\mathcal{L}_b(\Bx_b;\theta)$ in \eqref{eq: boris_symb_lik} given by \cite{beranger2018new}.

\begin{center}
    \begin{table}[ht!]
\footnotesize
\centering
  \setlength{\tabcolsep}{4pt}
\begin{tabular}{@{} r*{1}r*{11}{c} @{}}
\toprule
  & \multicolumn{5}{c}{$m=20$} & & \multicolumn{5}{c}{$m=50$} \\
  \cmidrule(lr){3-7}\cmidrule(lr){9-13}
 $\rho$ &$n$ & 5 & 10 & 100 & 1,000 & 100,000 &  & 5 & 10 & 100 & 1,000 & 100,000 \\ 
  \midrule
\multirow{4}{*}{0.0}  & $L_O$  & -0.001 & 0.015 & 0.006 & -0.003 & 0.000 &  & -0.009 & 0.001 & -0.001 & 0.011 & 0.000 \\ 
   &  & (0.126) & (0.123) & (0.146) & (0.068) & (0.004) &  & (0.087) & (0.082) & (0.100) & (0.108) & (0.004) \\ 
   & $L_{B_1}$ & -0.006 & 0.014 & 0.000 & -0.006 & 0.000 &  & -0.017 & 0.001 & -0.001 & -0.003 & 0.000 \\ 
   &  & (0.108) & (0.073) & (0.046) & (0.037) & (0.025) &  & (0.071) & (0.045) & (0.026) & (0.023) & (0.017) \\ 
   \cmidrule(lr){2-13}
 \multirow{4}{*}{0.3} & $L_O$ & 0.304 & 0.297 & 0.273 & 0.168 & 0.012 &  & 0.306 & 0.303 & 0.289 & 0.249 & 0.045 \\
   &  & (0.112) & (0.129) & (0.160) & (0.217) & (0.087) &  & (0.067) & (0.066) & (0.100) & (0.152) & (0.143) \\ 
   & $L_{B_1}$ & 0.306 & 0.307 & 0.299 & 0.306 & 0.299 &  & 0.308 & 0.306 & 0.301 & 0.305 & 0.300 \\ 
   &  & (0.102) & (0.067) & (0.038) & (0.032) & (0.023) &  & (0.058) & (0.043) & (0.028) & (0.018) & (0.014) \\ 
    \cmidrule(lr){2-13}
  \multirow{4}{*}{0.5} & $L_O$ & 0.505 & 0.499 & 0.490 & 0.426 & 0.212 &  & 0.509 & 0.503 & 0.494 & 0.488 & 0.315 \\ 
   &  & (0.094) & (0.105) & (0.134) & (0.204) & (0.298) &  & (0.058) & (0.055) & (0.083) & (0.076) & (0.274) \\ 
   & $L_{B_1}$ & 0.504 & 0.505 & 0.499 & 0.505 & 0.503 &  & 0.506 & 0.505 & 0.501 & 0.502 & 0.503 \\ 
   &  & (0.084) & (0.059) & (0.035) & (0.029) & (0.021) &  & (0.048) & (0.036) & (0.023) & (0.018) & (0.014) \\ 
   \cmidrule(lr){2-13}
  \multirow{4}{*}{0.7} & $L_O$ & 0.701 & 0.700 & 0.696 & 0.692 & 0.641 &  & 0.706 & 0.702 & 0.701 & 0.701 & 0.695 \\ 
   &  & (0.077) & (0.074) & (0.079) & (0.081) & (0.233) &  & (0.044) & (0.039) & (0.047) & (0.045) & (0.055) \\ 
  & $L_{B_1}$ & 0.702 & 0.705 & 0.700 & 0.703 & 0.703 &  & 0.704 & 0.704 & 0.700 & 0.701 & 0.703 \\ 
  &  & (0.060) & (0.043) & (0.029) & (0.021) & (0.018) &  & (0.034) & (0.025) & (0.019) & (0.015) & (0.012) \\
  \cmidrule(lr){2-13}
  \multirow{4}{*}{0.9} & $L_O$ & 0.901 & 0.900 & 0.901 & 0.901 & 0.903 &  & 0.902 & 0.901 & 0.901 & 0.900 & 0.902 \\ 
   &  & (0.030) & (0.026) & (0.025) & (0.028) & (0.023) &  & (0.017) & (0.014) & (0.016) & (0.016) & (0.015) \\ 
   & $L_{B_1}$ & 0.900 & 0.901 & 0.901 & 0.901 & 0.903 &  & 0.901 & 0.900 & 0.901 & 0.900 & 0.902 \\ 
  &  & (0.023) & (0.018) & (0.016) & (0.012) & (0.008) &  & (0.013) & (0.011) & (0.009) & (0.008) & (0.007) \\ 
   \bottomrule
\end{tabular}
\bigskip
\caption{\small Mean (and standard deviation) of the maximum likelihood estimate of the correlation $\rho$ over 100 replicate datasets. Estimates are based on the original construction of \cite{beranger2018new}, $L_O$, and our reworked construction, $L_{B_1}$. The table shows the number of random rectangles ($m$) in the likelihood, the number of data points per rectangle ($n$), and the true correlation $\rho$ between the two variables.
}   
\label{table: bvn}
\end{table}
\end{center}

\subsection{Example 2: Comparing the exact and approximate estimators}

The exact and approximate methods of Sections \ref{sec: exact method} and \ref{sec: approx method} involve a two-step procedure to estimate the likelihood. We now compare the results of the methods for each step in this process.
 We first compare the performance of path sampling (exact method) and the Taylor approximation (approximate method) in estimating the logarithm of the  likelihood function. We then examine the performance of the (exact) Poisson estimator and the (approximate) approximately bias-corrected estimator for estimating the likelihood function.

As an illustrating example, we choose $g_{X}(x;\mu_d,\Sigma_d)$ to be a $d$-dimensional multivariate normal distribution with $\mu_d =  0_d, \Sigma_d = 0.5I_d + 0.51_d 1_d^\top$, where $0_d$,  $1_d$ are respectively $d$-dimensional vectors of 0's and 1's and $I_d$ is a $d$-dimensional identity matrix. The integration region over the `random' rectangle is fixed as $B_1 = [-2,2]^d$,
for $d = 2,\dots,10$, and the number of observations in the rectangle is $n = 100$.

Table \ref{tab:ex2_p1} shows the mean and variance of the estimated log-likelihood $\widehat{\log L}$ for $\log L_{B_1} = n \log \int_{B_1} g_{X}(z;\mu_d,\Sigma_d)dz$, under both path sampling and the Taylor approximation methods taken over 1{,}000 replicated calculations, and for the given values of $\mu_d$ and $\Sigma_d$. Both methods provide similar results with the difference in estimates becoming larger as the number of dimensions increases. The largest difference is  0.624 for $d = 10$. The path sampler gives an unbiased estimator of the log-likelihood function (up to trapezoidal rule error), 
subject to an adequate integration over the temperature ladder. In contrast, the estimator obtained by the Taylor expansion is approximate, with the error increasing with dimension, but it has a lower variance than the path sampler. While the bias increases with dimension, this is not a major limitation in practice in the present work, since symbolic data analysis in its current form is primarily applied to problems with low-dimensional data. On the other hand, the computing time over 1{,}000 replications shows that the Taylor approximation only takes around 1\% of the path sampler's time. We conclude that, compared to the path sampler, the Taylor approximation has good accuracy and a significantly lower computing time, especially for lower dimensions. 

  \begin{table}[ht!]
    \setlength{\tabcolsep}{6pt}
    \centering
    \begin{tabular}{@{} r*{7}{c} @{}}
    \toprule
      {$d$} & \multicolumn{2}{c}{Mean of $\widehat{\log L}$} & \multicolumn{2}{c}{Var of $\widehat{\log L}$} & \multicolumn{2}{c}{Time (secs)} \\
      & path & Taylor & path & Taylor & path & Taylor\\
  \cmidrule(lr){2-3}\cmidrule(lr){4-5}\cmidrule(lr){6-7}
2 & \phantom{1}-8.649 & \phantom{1}-8.651 & 0.109 & 0.003 & 212.851 & 1.816 \\
3 & -12.226 & -12.209 & 0.187 & 0.011 & 284.973 & 2.489 \\
4 & -15.463 & -15.466 & 0.274 & 0.025 & 361.829 & 2.910 \\
5 & -18.462 & -18.497 & 0.405 & 0.047 & 426.571 & 3.372 \\
6 & -21.260 & -21.340 & 0.566 & 0.071 & 556.105 & 6.626 \\
7 & -23.890 & -24.013 & 0.680 & 0.097 & 587.027 & 4.385 \\
8 & -26.338 & -26.618 & 0.724 & 0.123 & 656.189 & 5.007 \\
9 & -28.720 & -29.105 & 0.850 & 0.171 & 670.050 & 5.785 \\
10 & -30.955 & -31.579 & 1.045 & 0.212 & 715.157 & 6.847\\
\bottomrule
    \end{tabular}
    \bigskip
        \caption{\small Mean, variance and execution time of $1{,}000$ replicate estimates of $\log L_{B_1} = n \log \int_{B_1} g_{X}(z;\mu_d,\Sigma_d) dz $ with $ n = 100$, where $g_{X}(z;\mu_d,\Sigma_d)$ is a $d$-dimensional multivariate normal distribution with $\mu_d = 0_d, \Sigma_d = 0.5 I_d + 0.5 1_d 1_d^\top, B_1 = [-2,2]^d$ and $d = 2,\dots,10$. For the path sampler (path), the temperature ladder is defined as $(t/T)^5$ with $T = 100$ and $t = 1,\dots,T$. The number of Monte Carlo draws at each temperature is $M=2{,}000$. For the Taylor approximation (Taylor), we also set $M = 2{,}000$. }\label{tab:ex2_p1} 
  \end{table}
  
\begin{table}[ht!]
    \centering
    \setlength{\tabcolsep}{6pt}
    \begin{tabular}{@{}
    r*{7}{c}
    @{}}
    \toprule
      {$d$} & \multicolumn{2}{c}{Mean  of $\log |\widehat{L}|$} & \multicolumn{2}{c}{Var of $\log |\widehat{L}|$} & \multicolumn{2}{c}{Time (secs)} \\
      & pois & bc & pois & bc & pois & bc \\
  \cmidrule(lr){2-3}\cmidrule(lr){4-5}\cmidrule(lr){6-7}
2 & \phantom{1}-8.657 & \phantom{1}-8.653 & 0.005 & 0.003 & \phantom{1}5.770 & 1.919 \\
3 & -12.256 & -12.217 & 0.077 & 0.011 &\phantom{1}7.433 & 2.450 \\
4 & -15.586 & -15.475 & 0.279 & 0.026 & \phantom{1}9.294 & 2.955 \\
5 & -18.788 & -18.512 & 0.510 & 0.043 & 10.483 & 3.476 \\
6 & -21.743 & -21.364 & 0.832 & 0.069 & 12.360 & 3.986 \\
7 & -24.634 & -24.094 & 0.980 & 0.102 & 13.618 & 4.443 \\
8 & -27.380 & -26.702 & 1.048 & 0.132 & 14.875 & 4.905 \\
9 & -29.958 & -29.236 & 1.189 & 0.162 & 16.810 & 5.748 \\
10 & -32.353 & -31.675 & 1.350 & 0.215 & 19.227 & 6.353 \\
\bottomrule
    \end{tabular}
    \bigskip
    \caption{\small Mean, variance and execution time of $1{,}000$ replicate estimates of the (absolute) value of the likelihood function $L_{B_1} = [\int_{B_1} g_{X}(z;\mu_d,\Sigma_d) dz]^{n}$, $n=100$, where the estimator of the log-likelihood function is provided by the approximate method. Results are presented on the log scale. For the Poisson estimator (pois), $\lambda = 3$, $a = n \gamma^d \sum_{i=1}^{d}\log \left((\Phi(u^{(i)},\mu_i,\Sigma_{ii}) -\Phi(l^{(i)},\mu_i,\Sigma_{ii})\right)-\lambda$, with $\gamma=0.97$, where $\Phi(.)$ denotes the cdf of a normal distribution. The abbreviation ``bc'' denotes the bias-corrected estimator.} \label{tab:ex2_p2}
\end{table}

Table~\ref{tab:ex2_p2} shows the mean and variance of the logarithm of the absolute likelihood estimator obtained by the Poisson method and the bias-correction method based on the approximate log-likelihood estimator $A_T$ and using identical settings to  Table \ref{tab:ex2_p1}.  We compare the results on the logarithmic scale as the target of interest, $L_{B_1}=[\int_{B_1} g_{X}(z;\mu_d,\Sigma_d)dz]^n$, is numerically close to zero for large values of $n$. Note that this is different from the estimator of the log-likelihood in Table \ref{tab:ex2_p1} ($\log\widehat{L} \neq \widehat{\log L}$). Both estimators provide similar results for the mean values across all dimensions, with the difference again increasing with $d$. 
The exact Poisson estimator has around 6--10 times the variance of that of the bias-corrected method, with around 3 times longer computing time.

In summary, we conclude that the estimators in the approximate method (the Taylor expansion with bias-correction) offers results close to those in the exact method (path sampling with Poisson estimator), but with significantly less computing time. Accordingly, we use the approximate method in the following analyses. However, for analyses that require exactness (defined as simulation-consistent estimates of functions of parameters with respect to the symbolic posterior distribution), the exact method can be used, albeit with some computational cost.

\subsection{Example 3: Implementation on a factor model}

We now explore the efficiency and accuracy of the methodology in a Bayesian factor model analysis, where observations are drawn from a multivariate normal $y_i\sim N_d(\cdot,\Sigma)$ distribution with the covariance matrix constructed via a low-rank approximation for efficient parameterisation. 
For $d$-dimensional observations, the covariance matrix 
is $\Sigma = LL^\top + D$, where $L$ is a lower triangular $d\times k$ matrix ($k \ll d$) and $D$ is a $d \times d$ diagonal matrix with positive entries. The factor model is then $y_i = \mu + L f_i + \epsilon_i$, where $f_i \sim N(0_k, I_k)$, $\epsilon_i \overset{\mathrm{iid}}{\sim}  N_d(0_d, D)$. 

To ensure a fair comparison between a standard Bayesian analysis using the micro-data and standard likelihood function, and that using the symbols and symbolic likelihood function, we use a Metropolis-within-Gibbs MCMC sampler for both methods, in which the elements in $\mu$, $L$ and $D$ are block updated conditioning on the other parameters. This avoids comparisons with samplers like e.g.~\cite{GewekeZhou2015} in which a number of latent variables, proportional to the number of data points $n$, are introduced to induce full conditionals with closed form, as this will negatively impact the efficiency of the standard likelihood approach for large $n$.

We generate datasets of size $n=$ 500{,}000, 1{,}000{,}000, and 5{,}000{,}000 independent observations from $y_i \sim N_d(\mu,\Sigma)$ for each of $d=3,\ldots,10$, and set $k=1$. The elements of the $d$-dimensional vector $\mu$ are equally spaced from 
$-1$ to 1. The covariance matrix is constructed with $\Sigma = LL^\top +D,$ where $\log D_{ii} \sim \mathrm{Uniform}(0.25,0.5)$, $L_{ij} \sim \mathrm{Uniform} (-0.5,0.5)$ for $i=1,\dots,d$ and $1\leq j \leq i $. 
A single random rectangle is constructed for each dataset using $q=0.005$, so that $B_1$ contains all micro-data lying within the central 99\% of each margin, and the remaining observations enter the likelihood $L(\Bx_e;\theta)$ in the standard way. (Section \ref{sec: Empirical study} explores the performance for varying $q$.) In the pseudo-marginal MCMC algorithm using the approximate method, we set $u$ as a collection of $M= 10{,}000 $ random numbers to estimate the likelihood function. To implement the signed block PMMH algorithm (Algorithm \ref{algo: PM-MCMC_SDA}), only one element in $u$ is updated randomly per MCMC iteration. For all analyses, we run the MCMC for $10{,}000$ iterations and use the last $5{,}000$ samples to estimate the posterior.

Table~\ref{tab:ex3} shows the average  mean squared error (AMSE) over all parameters ($\mu$, $\log D$ and $B$), and the sampler running time based on 100 replicate analyses. The AMSE is computed as
$\mathrm{AMSE}(\widehat{\theta},\theta) = \frac{1}{p}\sum_{i=1}^p (\widehat{\theta}_i-\theta_i)^2$,
where $\theta$ is the true value of a parameter with $p$ elements, $\{\theta_i\}$, and $\widehat{\theta}$ is its estimated posterior mean. Because of how the random hyper-rectangle $B_1$ is constructed, the proportion of external micro-data points is approximately $n_e/n\approx 1-(1-2q)^d$, which increases as $d$ increases.

From Table \ref{tab:ex3}, symbolic data analysis consistently yields a higher AMSE than the full-data approach, reflecting the loss of information induced by aggregation. Across all scenarios considered, SDA achieves substantial reductions in computation time relative to full-data inference, with speed-ups increasing considerably as the number of observations grows. However, when performance is assessed using a combined mean squared error criterion, the symbolic approach is most favourable in low-dimensional settings, where the substantial reduction in computation time outweighs the loss in statistical efficiency. As the dimension increases (for a fixed number of observations), the mean-squared error grows faster than the corresponding reduction in computation time, so that the overall mean squared error–time product becomes less favourable under this criterion and full-data inference is preferable when it is computationally feasible. Importantly, the primary motivation for symbolic data analysis is precisely to enable inference in settings where full-data likelihood evaluation is computationally or practically infeasible. This behaviour is largely attributable to the marginal construction of the symbols, which becomes increasingly coarse in higher dimensions and does not fully capture multivariate dependence. Developing more informative symbolic representations that better preserve dependence structure, such as multiple random rectangles, is left for future research.

Finally, while the absolute runtimes reported here are moderate, the computational gains become particularly valuable in settings requiring repeated analyses, large-scale simulation studies, or sequential updating, where full-data inference would be prohibitively expensive.

\begin{table}[ht!]
    \centering
    \setlength{\tabcolsep}{6pt}
    \renewcommand{\arraystretch}{0.5} % this reduces the vertical spacing between rows
\linespread{2}\selectfont\centering
    \begin{tabular}{@{} r*{1}r*{9}{c} @{}}
    \toprule
    &\multicolumn{9}{c}{n = 500{,}000}\\
    \midrule
      {dim} & $n_e/n(\%)$& \multicolumn{3}{c}{AMSE }  & \multicolumn{3}{c}{Time (secs)}  & Overall Ratio\\
      & & Full & {SDA} & {Ratio} & Full & {SDA} & {Ratio} & {Full/SDA} & \\
        \cmidrule(lr){3-5}\cmidrule(lr){6-8}
3 & 2.97 & 4.97$\times10^{-4}$ & 1.49$\times10^{-3}$ & 0.33 & 1027.19 & 199.31 & 4.71 & 1.57 \\
4 & 3.95 & 4.71$\times10^{-5}$ & 4.37$\times10^{-4}$ & 0.11 & 1216.31 & 263.79 & 4.26 & 0.46 \\
5 & 4.89 & 1.79$\times10^{-5}$ & 9.71$\times10^{-5}$ & 0.18 & 1421.60 & 328.23 & 4.02 & 0.74 \\
6 & 5.84 & 1.13$\times10^{-5}$ & 7.13$\times10^{-5}$ & 0.16 & 1765.21 & 401.37 & 4.02 & 0.64 \\
7 & 6.77 & 7.86$\times10^{-6}$ & 5.33$\times10^{-5}$ & 0.15 & 2030.02 & 485.10 & 3.77 & 0.56 \\
8 & 7.67 & 7.02$\times10^{-6}$ & 5.25$\times10^{-5}$ & 0.13 & 2116.21 & 553.95 & 3.62 & 0.48 \\
9 & 8.63 & 6.59$\times10^{-6}$ & 5.36$\times10^{-5}$ & 0.12 & 2483.95 & 635.73 & 3.46 & 0.43 \\
10 & 9.54 & 6.29$\times10^{-6}$ & 6.40$\times10^{-5}$ & 0.10 & 2707.51 & 711.10 & 3.39 & 0.33 \\
\midrule
 &\multicolumn{9}{c}{n = 1{,}000{,}000}\\
    \midrule
     {dim} & $n_e/n(\%)$& \multicolumn{3}{c}{AMSE }  & \multicolumn{3}{c}{Time (secs)}  & Overall Ratio\\
      & & Full & {SDA} & {Ratio} & Full & {SDA} & {Ratio} & {Full/SDA} & \\
        \cmidrule(lr){3-5}\cmidrule(lr){6-8}
3 & 2.97 & 3.36$\times10^{-4}$ & 7.74$\times10^{-4}$ & 0.43 & 1602.68 & 192.44 & 9.75 & 4.23 \\
4 & 3.94 & 3.03$\times10^{-5}$ & 2.50$\times10^{-4}$ & 0.12 & 2131.97 & 260.96 & 9.05 & 1.10 \\
5 & 4.89 & 6.38$\times10^{-6}$ & 5.58$\times10^{-5}$ & 0.11 & 2929.83 & 337.40 & 9.12 & 1.04 \\
6 & 5.85 & 5.82$\times10^{-6}$ & 5.65$\times10^{-5}$ & 0.10 & 3629.13 & 422.42 & 8.55 & 0.88 \\
7 & 6.78 & 4.08$\times10^{-6}$ & 5.91$\times10^{-5}$ & 0.07 & 4194.31 & 513.99 & 8.03 & 0.55 \\
8 & 7.72 & 3.67$\times10^{-6}$ & 7.51$\times10^{-5}$ & 0.05 & 5665.62 & 646.85 & 7.68 & 0.38 \\
9 & 8.63 & 3.04$\times10^{-6}$ & 9.99$\times10^{-5}$ & 0.03 & 4601.20 & 740.69 & 6.80 & 0.21 \\
10 & 9.56 & 3.33$\times10^{-6}$ & 9.79$\times10^{-5}$ & 0.03 & 5076.95 & 838.99 & 6.64 & 0.23 \\
\midrule
    &\multicolumn{9}{c}{n = 500{,}000{,}000}\\
    \midrule
     {dim} & $n_e/n(\%)$& \multicolumn{3}{c}{AMSE }  & \multicolumn{3}{c}{Time (secs)}  & Overall Ratio\\
      & & Full & {SDA} & {Ratio} & Full & {SDA} & {Ratio} & {Full/SDA} & \\
        \cmidrule(lr){3-5}\cmidrule(lr){6-8}
3 & 2.97 & 8.11$\times10^{-5}$ & 3.99$\times10^{-4}$ & 0.20 & 11114.85 & 422.05 & 27.11 & 5.50 \\
4 & 3.94 & 8.25$\times10^{-6}$ & 1.85$\times10^{-4}$ & 0.04 & 13631.53 & 725.75 & 18.80 & 0.84 \\
5 & 4.89 & 2.12$\times10^{-6}$ & 2.52$\times10^{-4}$ & 0.01 & 17534.34 & 1003.35 & 15.96 & 0.13 \\
6 & 5.85 & 1.00$\times10^{-6}$ & 3.15$\times10^{-4}$ & $<0.01$ & 19386.45 & 1264.63 & 14.23 & 0.05 \\
7 & 6.79 & 8.58$\times10^{-7}$ & 3.92$\times10^{-4}$ & $<0.01$ & 22119.67 & 1676.03 & 12.10 & 0.03 \\
8 & 7.71 & 7.24$\times10^{-7}$ & 4.12$\times10^{-4}$ & $<0.01$ & 20877.13 & 2016.66 & 11.39 & 0.02 \\
9 & 8.63 & 6.77$\times10^{-7}$ & 5.03$\times10^{-4}$ & $<0.01$ & 23493.89 & 2485.89 & 10.13 & 0.01 \\
10 & 9.53 & 6.53$\times10^{-7}$ & 5.65$\times10^{-4}$ & $<0.01$ & 25705.74 & 3032.59 & 9.25 & 0.01 \\

\bottomrule
    \end{tabular}
    \bigskip
    \caption{\small Average mean squared error (AMSE) Mean and execution time for 100 independent replications of the full and SDA-based analyses under datasets of size $n=$500{,}000, 1{,}000{,}000 and 5{,}000{,}000. The columns show the data dimension (dim), mean proportion of data external to the random rectangle ($n_e/n$), 
    AMSE of all parameters, and computing time (Time). The final column shows the ratio of the product of AMSE and computing time between the full-data approach versus SDA (Overall Ratio). } \label{tab:ex3}
\end{table}

\section{Empirical study: 2015 U.S.~domestic flight delays}

\label{sec: Empirical study}

We now explore a real-data application using a Bayesian linear model.
We analyse 2015 flight delay data, available from Kaggle ({\small \url{https://www.kaggle.com/usdot/flight-delays}}), provided by the U.S.~Department of Transportation's Bureau of Transportation Statistics. 
Ignoring cancelled flights, it comprises 5{,}714{,}008 on-time performance records of domestic flights operated by 14 large air carriers. The aim is to predict a flight's arrival delay given any departure delay, the original scheduled length of the flight, and the airline. Exploratory analysis of data subsets suggests that arrival delay variability increases with the log of the scheduled flight length.
One simple model is:
\begin{align*}
    y_{ij}  &= \beta_{0,i} + \beta_1 x_{1ij} + \beta_2 x_{2ij} + \epsilon_{ij},\qquad
    \epsilon_{ij} \sim N(0,\sigma_{ij}^2), \\
    \log(\sigma_{ij}^2) &= \alpha_0 + \alpha_1 (x_{2ij} - \overline{x}_{2i\cdot}),
\end{align*}
where $y_{ij}$ is the arrival delay (1 unit = 5 minutes) for the $j$-th flight of air carrier $i=1,\ldots,14$, $x_{1ij}$ is the associated departure delay, $x_{2ij}$ is the log scheduled flight length (1 unit = $\log (5)$ minutes), and $\overline{x}_{2i\cdot}$ denotes the average value of $x_{2ij}$ in the $i$th group. For example, a flight with scheduled length of 205 minutes that departed 11 minutes early and arrived 22 minutes early would have $y = -22/5$, $x_1 = -11/5$, $x_2 = \log (205/5)$. For this analysis, interest is in the posterior distribution of $\theta = (\beta_{0,1},\dots, \beta_{0,14} , \beta_1, \beta_2, \alpha_0, \alpha_1)^\top$. We specify a prior for each element in $\theta$ as $N(0,10^2)$ for simplicity.

\begin{figure}[ht!]
    \centering
    \includegraphics[width=1.0\linewidth]{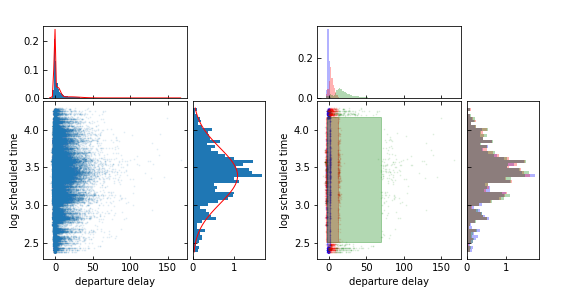}
    \caption{\small Demonstration of the random rectangle construction for Spirit Airlines ($i=8$), which has 115{,}193 observations. Left panel: The empirical relationship between departure delay  ($x_1$) and log of scheduled time ($x_2$). Top and right histograms show the marginal distributions, with density curves based on a fitted 3-component bivariate normal mixture on $(x_1,x_2)$.  Right panel: Three random rectangle summaries (one for each component),  obtained with $q = 0.01$. Histogram, rectangle, and micro-data colours indicate mixture component membership.}
    \label{fig:Ex_symb}
\end{figure}

The micro-data for aggregation are the triple $(y,x_1,x_2)$ (dropping $ij$ subscripts for ease of exposition). 
Because the symbolic likelihood function in \eqref{eq2: symb_lik} integrates over all aggregated data this means that, unlike a direct analysis on the micro-data which only requires the conditional model $y|x_1,x_2$, an analysis based on random rectangles requires a joint model for $(y,x_1,x_2)$. 
For large or complex datasets, assuming a simple parametric model (e.g.~a multivariate normal distribution, or a uniform distribution) is unlikely to be appropriate: see e.g.~Figure \ref{fig:Ex_symb} (left panel) for the detailed structure of the empirical distribution of $(x_1,x_2)$ for one of the airlines (Spirit Airlines).

Writing this joint model as  $g(y,x_1,x_2| \theta) = g(y| x_1,x_2,\theta) g(x_1,x_2)$, where $g(y|x_1,x_2,\theta)$ is the above regression model, an unbiased estimate of the integral in the symbolic likelihood \eqref{eq2: symb_lik} is given by the final expression in
  \begin{align}
    \int_{B_1} g(y,x_1,x_2|\theta) dx_1 dx_2 dy &= \int_{B_1} g(y| x_1,x_2,\theta) g(x_1,x_2) dx_1 dx_2 dy \nonumber \\
    &= \mathbb{E}_{(x_1,x_2)\sim g(x_1,x_2)} \bigg[\int_{B_1(y)} g(y| x_1, x_2, \theta) dy \bigg] \nonumber \\
    &\approx \frac{1}{M}\sum_{m=1}^M \bigg( \Phi(y_u| x_1^{(m)},x_2^{(m)},\theta) - \Phi(y_l| x_1^{(m)},x_2^{(m)},\theta)\bigg),
    \label{eq: ex1_intergral}
  \end{align}
where $B_1$ is a $d=3$ dimensional random rectangle over the space of $(y,x_1,x_2)$, $B_1(y)=(y_l,y_u)$ is the univariate marginal random interval of $B_1$ over $y$,
$(x_1^{(m)}, x_2^{(m)})$ are samples from $p(x_1,x_2)$, $m=1,\dots,M$, and $\Phi(\cdot)$ is the normal distribution function. Here it is the aggregation over the predictors $(x_1,x_2)$, rather than over $y$, that induces integral intractability.

\begin{figure}[ht!]
    \centering
    \includegraphics[width=1.0\linewidth]{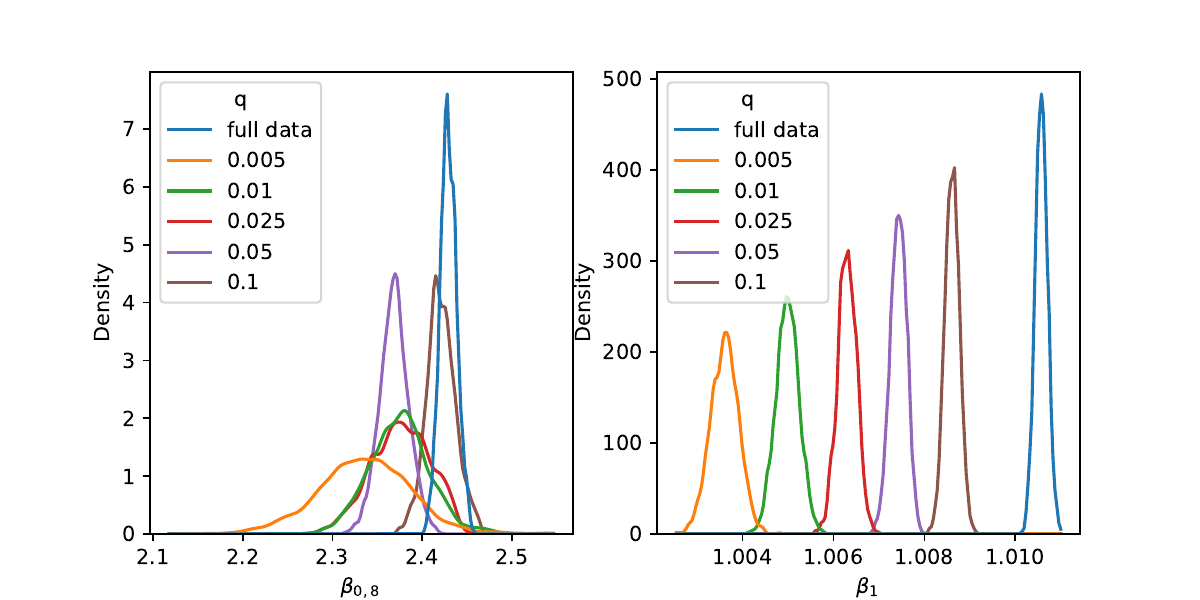}
    \caption{\small 
    Estimated marginal posterior densities of $\beta_{0,8}$ (8 corresponds to Spirit airline) and $\beta_1$ for different values of random rectangle specification, $q$.
    }
    \label{fig:Ex extra}
\end{figure}

We use a finite mixture of normal distributions to approximate $p(x_1,x_2)$, independently for each airline. For each air carrier $i$, we construct $K$ random rectangles $B_1$ over $(y,x_1,x_2)$ as follows:
\begin{enumerate}
 \item   Fit a $K$-component mixture of normals distribution for $(x_{1ij},x_{2ij})$, $j = 1,\dots, n_i$. The number of components, $K$, is chosen as the smallest value such that the Bayesian information criteria (BIC) score doesn't strongly improve by adding one more component, while also ensuring that all components contain at least 10,000 observations (see step \ref{xxx} ). This number is justified below. For the 14 air carriers, $K= 4, 4, 5, 2, 5, 2, 2, 3, 2, 2, 2, 2, 2, 2$ components are selected. Larger numbers of components per airline will describe the data more accurately, at increased computational costs for inference.
 
 \item \label{xxx} 
Allocate each observation $(x_{1ij},x_{2ij})$ to its most likely mixture component. For each component, construct a random rectangle $B_1(x_1,x_2)$ over $(x_1,x_2)$ that contains those observations that are in the central $(1-2q)\%$ of data in each margin, for some $q\in[0,0.5]$. Extract those observations that do not fall inside the random rectangle.
\item For each component, identify the values of $y$ such that the associated $(x_1,x_2)$ values are in $B_1(x_1,x_2)$: $\{y: (x_1,x_2)\in B_1(x_1,x_2)\}$. Write the range of these $y$-values as $B_1(y)=(y_l,y_u)=(\min\{y\},\max\{y\})$. Construct the final random rectangle over $(y,x_1,x_2)$ as $B_1=B_1(y)\times B_1(x_1,x_2)$.
\end{enumerate}
The above process results in 39 random rectangles $B_1$ over $(y,x_1,x_2)$, one for each mixture component. Each of these has some proportion  ($1-(1-2q)^2$) of the observations allocated to the component retained as micro-data $\Bx_e$ external to the rectangle.
Figure \ref{fig:Ex_symb} illustrates the random rectangle construction process for observations from one airline (Spirit Airlines) with $115{,}193$ observations. The resulting hyper-rectangles (right panel) summarise 62{,}264 (orange), 33{,}867 (blue) and 14{,}974 (green) observations. There are 4{,}088 individual points ($\approx 3.5\%$) that remain external to any interval. While the fitted mixture model is not imperfect, the parameters estimated from the fitted model using these random rectangle summaries are close to those using the full dataset; see Table \ref{tab:empirical} and Figure \ref{fig:Ex extra}.

Inspection of \eqref{eq: ex1_intergral} suggests that evaluating the symbolic likelihood can be more computationally costly than the micro-data likelihood when the number of Monte Carlo draws $M$ is greater than $(n-n_b-n_e)/2$ (loosely equating the time to evaluate a density function and a distribution function). For the implemented (approximate method) block pseudo-marginal MCMC sampler, a low enough variance (so that the pseudo-marginal chain avoids getting stuck) is achieved with $M\approx4{,}000$. For this reason, we set the minimum number of observations within each random rectangle to be slightly more than double this value, at $10{,}000$ observations.

Based on an MCMC sampler of length 20,000, with half of this discarded as burn-in, the performance of the resulting symbolic posterior distribution is evaluated through the following measures: the mean absolute percentage error, $\mathrm{MAPE}( \widehat{\theta}_f,\widehat{\theta}_s) = \frac{1}{p} \sum_{i=1}^p (|\widehat{\theta}_{f,i} - \widehat{\theta}_{s,i} |)/| \widehat{\theta}_{f,i}|$, and the root mean squared error, $\mathrm{RMSE}(\widehat{\theta}_f,\widehat{\theta}_s) = (\frac{1}{p} \sum_{i=1}^p (\widehat{\theta}_{f,i} - \widehat{\theta}_{s,i})^2)^{1/2}$, where $\widehat{\theta}_f$ and $\widehat{\theta}_s$ are the estimated posterior mean vectors under the full-data and SDA-based analyses, respectively.

Table \ref{tab:empirical} shows both estimated performance measures as a function of $q$, as well as the total computing time broken down by random rectangle construction time and MCMC sampler time. 
As the proportion of data summarised by the random rectangles ($q$) decreases, both RMSE and MAPE reduce as a larger proportion of micro-data is included in the analysis. However, at the same time, the computational overheads increase. E.g. a MAPE of 3\% can be achieved at half the computational speed of the full analysis. Whereas if a MAPE of 8\% can be tolerated, the computational speed up is over 12 times.
Figure \ref{fig:Ex extra} visualises the estimated posterior marginal distributions for $\beta_{0,8}$ (Spirit Airlines) and $\beta_1$ using different choices for $q$. For both parameters there is some bias and increased variance for lower $q$, both of which reduce as $q$ increases. This effect for $\beta_1$, estimated from the data from all 14 airlines, is likely negligible given the scale. 
%The estimated posterior for $\beta_{0,8}$ is practically useful for $q=0.1$.

\begin{table}[ht!]
    \centering
    \setlength{\tabcolsep}{6pt}
    %\begin{tabular}{@{}l*{7}{c}@{}}
    \begin{tabular}{lc|cc|lllc}
    \toprule
      $q$ & $n_e/n$ & RMSE & MAPE & \multicolumn{3}{c}{Time (secs)} & \\
      &  &  & & Prep &  MCMC & Total & Ratio\\
      \midrule
0.005 & 0.028 & 0.17 & 0.08 & 31.71 & \phantom{11}934.05 & \phantom{11}965.76 & \phantom{1}0.08 \\
0.01 & 0.038 & 0.15 & 0.07 & 39.48 & \phantom{1}1453.83 & \phantom{1}1493.31 & \phantom{1}0.13 \\
0.025 & 0.092 & 0.15 & 0.06 & 39.68 & \phantom{1}2163.29 & \phantom{1}2202.98 & \phantom{1}0.19 \\
0.05 & 0.176 & 0.12 & 0.05 & 39.82 & \phantom{1}3378.86 & \phantom{1}3418.68 & \phantom{1}0.30 \\
0.1 & 0.339 & 0.07 & 0.03 & 40.44 & \phantom{1}5664.21 & \phantom{1}5704.64 & \phantom{1}0.49 \\
\midrule
Full & 1.000 & -- & -- & \phantom{0}0.00 & 11529.74 & 11529.74 & \phantom{1}1.00 \\
\bottomrule
    \end{tabular}
    \bigskip
    \caption{\small Performance measures (RMSE, MAPE) and computation time of estimated posterior means based on $20{,}000$ MCMC iterations under the symbolic and full-data likelihoods, for different random rectangle specification $q$. $n_e/n$ indicates the proportion of data not summarised in random rectangles. `Prep' denotes the time for constructing the random rectangles, and `MCMC' the time to run the sampler. Ratio denotes the total time of the symbolic versus full-data analyses, with values $<1$ indicating shorter computation times for the SDA analysis.
    } \label{tab:empirical}
\end{table}

%%%%%%%%%%%%%%%%%%%%%%%%%%%%%%%%%%%%%%%%%%%%%%%%%%%%%%%%%%%%%%%%%%%%%%%%%%%%%%%%%%%%%%%%%
%%%%%%%%%%%%%%%%%%%%%%%%%%%%%%%%%%%%%%%%%%%%%%%%%%%%%%%%%%%%%%%%%%%%%%%%%%%%%%%%%%%%%%%%%
\section{Conclusions and discussion}
%%%%%%%%%%%%%%%%%%%%%%%%%%%%%%%%%%%%%%%%%%%%%%%%%%%%%%%%%%%%%%%%%%%%%%%%%%%%%%%%%%%%%%%%%
%%%%%%%%%%%%%%%%%%%%%%%%%%%%%%%%%%%%%%%%%%%%%%%%%%%%%%%%%%%%%%%%%%%%%%%%%%%%%%%%%%%%%%%%%

\label{sec: conclusion and dis}

Our paper extends the class of models available through the symbolic likelihood-based approach of \cite{beranger2018new} to include those where the integral $p_b(\theta)=\int_{B_b} g_X(z;\theta)dz$ is unavailable in closed form (but where an unbiased estimate is available), via a pseudo-marginal MCMC approach. `Symbolic' methods provide one strategy for fitting models to large, complex datasets by summarising the data into descriptive distributions, including random hyper-rectangles and random histograms. This enables the likelihood function to be evaluated significantly more quickly than the standard likelihood function for the micro-data, although with some cost to analytical accuracy given the loss of information in the representation of the data. In addition, we have reworked the representation of min-max random rectangles from \cite{beranger2018new} to allow better representation and estimation of dependence between the summarised variables, resolving the previous issues of dependence parameter magnitude underestimation for large numbers of micro-data within random rectangles.

Within the pseudo-marginal framework we have developed two approaches to generate unbiased estimates of $p_b(\theta)^n$ when an unbiased estimate of $p_b(\theta)$ is available. The first of these combines path sampling with the Poisson estimator, and with the pseudo-marginal MCMC scheme targets a posterior based on the absolute value of the likelihood estimator. Through an importance sampling scheme (the signed PMMH algorithm), unbiased expectations with respect to the desired target posterior (in this case the symbolic posterior) can be achieved. This likelihood estimator is unbiased, although it has a large computational cost.
In contrast, the second approach is approximate, based on a Taylor expansion of the log estimator, and a bias-correction approach based on an assumption of normality of this estimator. However the computational overheads are far lower than for the exact estimator. In the simulation studies we performed, the resulting analysis approximations were minor, and were far outweighed by the increase in computational speed.

While we have focused on random rectangle micro-data summaries, the ideas and algorithms developed here trivially extend to random histogram micro-data summaries which retain more information about the micro-data than a single random rectangle, as the resulting likelihood functions are of the form $\prod_b p_b(\theta)^{n_b}$. In this case, the distributional summary design question relates more to the size, location and number of the histogram bins, rather than the proportion of micro-data, $q\approx n_e/n$, retained outside of the random rectangle. Principled methods for the construction of distributional summaries remain the subject of active research that we leave to future work.

We note that there are some limitations on the scalability of SDA methods as the dimension of the data space, $d$, increases. 
Mechanically, if the density $g_X$ has a tractable distribution function, then the integral $p_b(\theta)$
%in the symbolic likelihood function \eqref{eq2: symb_lik} 
has $2^d$ terms via the inclusion-exclusion principle, for each random rectangle considered. This is one of the motivations for the current work: at some point it is more efficient to estimate the integral computationally, regardless of the availability of the distribution function. In combination with the composite-likelihood approach of \cite{whitaker2020composite}, which reduced the need to work with $d$-dimensional histograms down to much lower-dimensional marginal histograms, this provides a route to practically extend the methods developed here to higher-dimensional data.
More conceptually, similar to the construction of histograms, the location of data within any bin becomes less certain as $d$ increases and the data become increasingly sparse. The constructed bin contains limited information about the dependence across dimensions. As a consequence, to maintain inferential accuracy, the volume of the bin should intuitively decrease as dimension increases, and the number of bins also increase to cover the data. More generally, methods for constructing bins that efficiently cover sparse data remains an open research question. Recent work on random partitions  \citep{fan2025dataRBP,fan2018RBP} may offer some ideas in this regard.

Finally, optimal tuning strategies for pseudo-marginal algorithms are developed in a number of papers \citep[e.g.]{pitt2012some, doucet2015efficient,sherlock2015efficiency, quiroz2021block, yang2022correlated}. Developing such guidelines for our estimators is a topic for future work.

\section*{Acknowledgements}
We thank the Associate Editor and two referees for helpful comments that significantly improved the manuscript. YY was financially supported by a University International Postgraduate Award from UNSW Sydney. MQ is partially supported by the Marine Ecosystems Research Mobilising AI and Data (MERMAID) Collaboration (CLB-3127). SAS and BB are supported by the Australian Research Council (FT170100079 and DP220103269).

\bibliographystyle{apalike}  
\bibliography{references}

\appendix
\renewcommand \appendixpagename{Appendix}
\appendixpage

%%%%%%%%%%%%%%%%%%%%%%%%%%%%%%%%%%%%%%%%%%%%%%%%%%%%%%%%%%%%%%%%%%%%%%%%%%%%%%%%%%%%%%%%%
%%%%%%%%%%%%%%%%%%%%%%%%%%%%%%%%%%%%%%%%%%%%%%%%%%%%%%%%%%%%%%%%%%%%%%%%%%%%%%%%%%%%%%%%%
\section{Path sampler details}
%%%%%%%%%%%%%%%%%%%%%%%%%%%%%%%%%%%%%%%%%%%%%%%%%%%%%%%%%%%%%%%%%%%%%%%%%%%%%%%%%%%%%%%%%
%%%%%%%%%%%%%%%%%%%%%%%%%%%%%%%%%%%%%%%%%%%%%%%%%%%%%%%%%%%%%%%%%%%%%%%%%%%%%%%%%%%%%%%%%
\label{app: ps}

Let $0\leq t \leq 1$, and denote $h_t(z; \theta) = g_{X}(z; 
\theta)^t, z \in B_1$, $B_1 \displaystyle \subseteq \mathbb{R}^d$; then the following equation holds
\begin{equation*} \label{eq: path_sampler_appendix}
    \log \int_{B_1} g_X(z;\theta) dz = 
    \int_0^1 \mathbb{E}_{q_t(z;\theta)} \left[ \frac{d}{dt} 
    \log h_t(z;\theta) \right]dt  + \log \int_{B_1} 1 dz.
\end{equation*}

\begin{proof}

It is straightforward to see that  $h_0(z; \theta) = 1$ and $h_1(z;\theta) = g_{X}(z;\theta)$.

Let $\phi_t(\theta) = \int_{B_1} h_t(z;\theta) dz $, then $\log(\phi_1(\theta)) = \log \left( \int_{B_1} g_{X}(z;\theta) dz \right)$, which is our target, and $\phi_0(\theta) = \int_{B_1} 1 dz =$ volume of $B_1$. Then,
\begin{align*}
    \frac{d}{dt} \log \phi_t(\theta) & = \frac{1}{\phi_t(\theta)} \frac{d}{dt} \phi_t(\theta) = \frac{1}{\phi_t(\theta)} \frac{d}{dt} \left( \int_{B_1} h_t(z;\theta)dz \right) \\
    & =  \frac{1}{\phi_t(\theta)} \int_{B_1} \frac{d}{dt} h_t(z;\theta ) dz \\
    & = \frac{1}{\phi_t(\theta)} \int_{B_1} h_t(z;\theta) \frac{d}{dt} \log h_t(z;\theta)dz\\
    &= \int_{B_1} \frac{h_t(z;\theta)}{\phi_t(\theta)} \frac{d}{dt} \log h_t(z;\theta) dz\\
    &=\int_S q_t(z;\theta) \frac{d}{dt} \log h_t(z;\theta)dz \quad \mbox{where } q_t(z;\theta)  = \dfrac{h_t(z;\theta)}{\phi_t(\theta)}\\
    &= \mathbb{E}_{q_t(z;\theta)} \left[ \frac{d}{dt} \log h_t(z;\theta)  \right].
\end{align*}
Integrating from 0 to 1,
\begin{align*}
    \left[ \log \phi_t(\theta) \right]^{1}_{0} &= \log \phi_1(\theta) - \log \phi_0(\theta)
    = \int_0^1 \mathbb{E}_{q_t(z;\theta)} \left[ \frac{d}{dt} \log h_t(z;\theta) \right] dt.
\end{align*}
\end{proof}
Note that when numerically estimating this integral in practice, the range of $t$ is taken to be $[\epsilon,1]$ for some small value $\epsilon>0$, e.g.~$\epsilon=(1/T)^5$ (Section \ref{sec: exact method}). This is because the variance of $z \sim q_t(z;\theta)$ depends on the reciprocal of $t$ and  $g_X(z;\theta)$ is a normal density function.

\section{Scaling of the path sampler variance with respect to the dimension of the data}\label{app:variance_scaling}

This section presents a heuristic derivation of the scaling of the variance of the estimator of the right-hand side of \eqref{eq: path_sampler}. It suffices to consider the estimator of \eqref{eq:integral}, since the other terms on the right-hand side of  \eqref{eq: path_sampler} do not depend on $d$ or are constant. 

First, note that $$\frac{d}{dt}\log h_t(z;\theta) = \frac{d}{dt} t \log g_X(z;\theta)=\log g_X(z;\theta),$$
so we can rewrite \eqref{eq:integral} as
\begin{align}\label{eq:integral_rewritten}
    I =\int_0^1 \mathbb{E}_{q_t(z;\theta)} \bigg[ \frac{d}{dt}\log h_t(z;\theta) \bigg]dt =\int_0^1 \mathbb{E}_{q_t(z;\theta)} \bigg[ \log g_X(z;\theta) \bigg]dt.  
\end{align}
We can estimate \eqref{eq:integral_rewritten} as
\begin{align}\label{eq:integral_rewritten_estimated}
    \widehat{I} = \sum_{i=1}^n w_i \widehat{\mathbb{E}}_{q_{t_i}(z;\theta)} \bigg[ \log g_X(z;\theta)\bigg],
\end{align}
where $w_i$ are the quadrature weights of the (one-dimensional) numerical integration (e.g.\ the trapezoidal method) based on $n$ grid points, and the Monte Carlo estimate of the expectation is
$$\widehat{\mathbb{E}}_{q_{t_i}(z;\theta)} \bigg[ \log g_X(z;\theta)\bigg] = \frac{1}{M} \sum_{m=1}^M \log g_X\left(z^{(t_i)}_m; \theta\right), \quad \text{with } z^{(t_i)}_m \sim q_{t_i}(z;\theta)\propto g_X(z;\theta)^{t_i}.$$
Assuming that the path sampling generates $M$ independent samples across the different temperatures, and that within a given temperature the $M$ samples are independent,  we obtain
\begin{align}\label{eq:integral_rewritten_estimated}
    \mathrm{Var}(\widehat{I}) = \sum_{i=1}^n w^2_i  \frac{1}{M}\mathrm{Var}_{q_{t_i}}\bigg[\log g_X(z;\theta)\bigg].
\end{align}
To compute the order of $\mathrm{Var}_{q_{t_i}}\bigg[\log g_X(z;\theta)\bigg]$, assume that $g_X(z;\theta)$ is a multivariate Gaussian with independent components, and that $$q_t(z;\theta) \propto \prod_{i=1}^d N(z_i|\mu_i,\sigma^2_i)^t \mathbbm{1}_{[a_d,b_d]}(z_i),\quad \text{with }  a_d = -c\sqrt{d}, b_d =+c\sqrt{d},  $$ 
where $N(\cdot | \mu, \sigma^2)$ denotes the univariate normal density with mean $\mu$ and variance $\sigma^2$, and $\mathbbm{1}_{A}(x)=1$ is the indicator function of the set $A$. The assumption that the truncation region (the box) grows with $d$ prevents the probability of the tempered Gaussian lying in the box from collapsing to zero as $d$ increases. This scaling is motivated by the fact that if $z\sim N(\mu, I_d)$ then $$(z-\mu)^\top(z-\mu) \sim \chi^2(d),$$ which has mean $d$ and variance $2d$. Consequently, $0.5(z-\mu)^\top(z-\mu)$ concentrates around $d/2$ with standard deviation of order $\sqrt{d}$.

Under the above assumptions, the log-density of $g_X(z;\theta)$ decomposes as
$$\log g_X(z;\theta) = -\frac{d}{2}\log(2\pi) - \frac{1}{2} \sum_{i=1}^d\log \sigma^2_i - \frac{1}{2}\sum_{i=1}^d X_i, \quad X_i = \frac{(z_i - \mu_i)^2}{\sigma^2_i}, \,\, z_i \sim  q_t(z;\theta).$$
Because $q_t(z;\theta)$ factorises across coordinates by assumption, the $X_1,\dots, X_d$ remain independent under truncation. Therefore, the variance simplifies to a sum,
\begin{align}\label{eq:sum_of_individual_variances}
  \mathrm{Var}_{q_t}\left[\log g_X(z;\theta)\right] & = \frac{1}{4}\sum_{i=1}^d \mathrm{Var}_{q_t}\left( X_i\right).  
\end{align}
We now show that each coordinate contributes an independent variance term of order $\mathcal{O}(1)$, and thus the total variance across $d$ components scales linearly with $d$. To derive the order of $\mathrm{Var}_{q_t}\left( X_i\right)=\mathbb{E}_{q_t}\left( X^2_i\right) - \mathbb{E}_{q_t}\left( X_i\right)^2$, 
note that for the truncated density (conditional on $-c\sqrt{d}\leq z_i \leq c\sqrt{d}$), 
\begin{align}\label{eq:truncated_moments}
\mathbb{E}_{q_t}\left(X^k_i\right) & =\frac{\mathbb{E}_{\widetilde{q}_t}\left(X^k_i\mathbbm{1}(-c\sqrt{d} \leq z_i \leq c\sqrt{d})\right)}{\Pr_{\widetilde{q}_t}\left(|z_i| \leq c\sqrt{d}\right)},
\end{align}
where $\widetilde{q}_t$ is the untruncated density,
\begin{align*}%\label{eq:q_no_truncation}
    \tilde{q}_t(z_i;\theta) = \frac{g_X(z_i)^t}{\int_{\mathbb{R}} g_X(z_i)^t dz_i} = N(z_i|\mu_i,\sigma^2_i/t).
\end{align*}
We now show that $\mathbb{E}_{q_t}(X^k_i)=\mathcal{O}(1)$ for $k=1,2$, which implies $\mathrm{Var}_{q_t}(X_i)=\mathcal{O}(1)$.
We can bound the truncated moments by applying Cauchy-Schwartz to the numerator in \eqref{eq:truncated_moments},
\begin{align}\label{eq:upper_bound_truncated_conditional_moment}
\mathbb{E}_{q_t}\left(X^k_i\right) & \leq \frac{\sqrt{\mathbb{E}_{\widetilde{q}_t}(X^{2k}_i)}\sqrt{\Pr_{\widetilde{q}_t}\left(|z_i| \leq c\sqrt{d}\right)}}{\Pr_{\widetilde{q}_t}\left(|z_i| \leq c\sqrt{d}\right)}= \frac{\sqrt{\mathbb{E}_{\widetilde{q}_t}(X^{2k}_i)}}{\sqrt{\Pr_{\widetilde{q}_t}\left(|z_i| \leq c\sqrt{d}\right)}}.    
\end{align}
The upper bound in \eqref{eq:upper_bound_truncated_conditional_moment} depends on the moments and probability under the untruncated $\widetilde{q}_t$, which are simpler to compute compared to those of the truncated $q_t$. For the moments, note that under $\widetilde{q}_t$, $z_i\sim N(\mu_i, \sigma_i^2/t)$, and since $$X_i = \frac{1}{t}\left(\frac{z_i - \mu_i}{\sigma_i/\sqrt{t}}\right)^2,$$
it follows that $X_i \sim (1/t) \chi ^2(1)$, i.e.\ a scaled $\chi^2(1)$ distribution with $$\mathbb{E}_{\widetilde{q}_t}(X_i)=1/t, \,\,  \mathbb{E}_{\widetilde{q}_t}(X_i^2)=3/t^2, \,\, \text{and } \mathbb{E}_{\widetilde{q}_t}(X_i^4)=105/t^4,$$ all bounded in $d$. Thus, the numerator of the upper bound in \eqref{eq:upper_bound_truncated_conditional_moment} is $\mathcal{O}(1)$. For the denominator, the sub-Gaussian tail bound (see e.g.\ \citealt[Eq (2.15)]{Vershynin2018highdim})
$$\Pr(|z_i| \geq s) \leq 2\exp\left(-Ks^2 \right), \quad \text{for some constant } K>0,$$
with $s=c\sqrt{d}$, gives $\Pr(|z_i| \geq c\sqrt{d}) \leq 2\exp\left(-Kc^2d \right)$ and thus
$$\textstyle \Pr_{\widetilde{q}_t}\left(|z_i| \leq c\sqrt{d}\right) = 1 - \Pr_{\widetilde{q}_t}\left(|z_i| \geq c\sqrt{d}\right) = 1 - o(1)=\mathcal{O}(1).$$
We conclude that the first two moments $\mathbb{E}_{q_t}(X_i)$ and $\mathbb{E}_{q_t}(X^2_i)$ are bounded in $d$, implying that $\mathrm{Var}_{q_t}(X_i)=\mathcal{O}(1)$ and thus, by $\eqref{eq:sum_of_individual_variances}$,
$$ \mathrm{Var}_{q_t}\left[\log g_X(z;\theta)\right] = \mathcal{O}(d).$$
Hence, under the assumptions above, the variance of \eqref{eq:integral_rewritten_estimated} is
$$    \mathrm{Var}(\widehat{I}) = \sum_{i=1}^n w^2_i  \frac{1}{M}\mathcal{O}(d) = \frac{\mathcal{O}(d)}{M}\sum_{i=1}^n w^2_i = \mathcal{O}\left(\frac{d}{M}\right),$$
since $w_i$ (univariate integral weights)  do not depend on $d$.

\section{Some properties of the Poisson estimator}

{\label{app: pois est}}

Recall that the Poisson estimator is  
\begin{equation*}
        \widehat{\exp(A_P)} =\exp(a+\lambda) \prod_{h=1}^{\chi} \dfrac{(\widehat{A}_P^{\,(h)} - a)}{\lambda},
\end{equation*}
where $\chi\sim\mbox{Poisson}(\lambda)$, and the $\widehat{A}_P^{(h)}$ are replicate estimators of $A$ such that $\mathbb{E}[\widehat{A}_P^{(h)}]=A$.
%, e.g.\ $\widehat{A}_P$ in Section \ref{sec: exact method}. 
The following properties of the Poisson estimator are given in \cite{papaspiliopoulos2011monte}. We state and prove them here with the notation used in our paper.

\begin{enumerate}
    \item $\mathbb{E}(\widehat{\exp(A_P)}) =\exp(A)$.
    \item  $\mathrm{Var}(\widehat{\exp(A_P)}) =\exp\left(  \dfrac{(A-a)^2}{\lambda}+\lambda + 2a + \dfrac{\widehat{\sigma}^2_A}{\lambda}  \right) -\exp(2A)$, where $\widehat{\sigma}^2_A=\mbox{Var}(\widehat{A}_P^{(h)})$.
    \item The optimal value of $a$ which minimises the variance of $\widehat{\exp(A_P)}$ is $a_{\mathrm{opt}} = A -\lambda$.
\end{enumerate}

We use the following results for the Poisson distribution in the proof. If $\chi \sim \text{Poisson} (\lambda)$ and $A<\infty$,
\begin{enumerate}
    \item[(i.)] $\mathbb{E}_\chi (A^\chi) =\exp [(A-1)\lambda]$.
    \item[(ii.)] $\mathrm{Var}_\chi (A^{\chi}) =\exp(-\lambda) [\exp(A^2\lambda) -\exp(2A\lambda -\lambda) ] $.
\end{enumerate}

\begin{proof}{Property 1}

Let $\widehat{A}_P=(\widehat{A}_P^{(1)},\ldots,\widehat{A}_P^{(\chi)})$ collect all estimators of $A$ used in the Poisson estimator. Then,
\begin{align*}
    \mathbb{E}(\widehat{\exp(A_P)}) &= \mathbb{E}_\chi \left[ \mathbb{E}_{\widehat{A}_P \mid \chi} \left[\exp(a+\lambda) \prod_{h=1}^{\chi} \frac{\widehat{A}_P^{\,(h)}-a}{\lambda} \right]\right] \\
    & =\exp(a+\lambda)  \mathbb{E}_\chi \left[  \left(\frac{A -a }{\lambda} \right)^\chi \right]  \\
    &=\exp(a+\lambda) +\exp\left( \frac{A-a}{\lambda}\lambda -\lambda\right) \quad \mbox{(using result (i.))}\\
    &=\exp(a +\lambda)\exp(A-a -\lambda) \\
    &=\exp(A),
\end{align*}
where $\mathbb{E}_{\widehat{A}_P \mid \chi}$ means taking the conditional (on $\chi$) expectation over (independent) $\widehat{A}_P^{(1)},\ldots,\widehat{A}_P^{(\chi)}$.
\end{proof}

We next derive the variance of $\widehat{\exp(A_P)}$ to obtain the optimal value of the lower bound $a$
\begin{proof}{Property 2}

\begin{align*}
    \mathrm{Var}(\widehat{\exp(A_P)}) &= \mathrm{Var}\left(\exp(a+\lambda) \prod_{h=1}^{\chi} \frac{\widehat{A}_P^{\,(h)}-a}{\lambda}\right)\\
    &=\exp(2a+2\lambda) \mathrm{Var}\left(\prod_{h=1}^{\chi} \frac{\widehat{A}_P^{\,(h)}-a}{\lambda}\right)\\
    &=\exp(2a + 2\lambda) \left(  \mathrm{Var}_\chi \mathbb{E}_{\widehat{A}_P\mid \chi}  \prod_{h=1}^{\chi} \frac{\widehat{A}_P^{\,(h)}-a}{\lambda} + \mathbb{E}_\chi \mathrm{Var}_{\widehat{A}_P\mid \chi} \prod_{h=1}^{\chi} \frac{\widehat{A}_P^{\,(h)}-a}{\lambda} \right)\\
    &=\exp(2a+2\lambda) (C+D),\\
    \text{with } C &=   \mathrm{Var}_\chi \mathbb{E}_{\widehat{A}_P\mid \chi}  \prod_{h=1}^{\chi} \frac{\widehat{A}_P^{\,(h)}-a}{\lambda}\\
    &=\exp(-\lambda) \left[ \exp \left(\left(\left[\frac{A-a}{\lambda}\right]^2\lambda\right) -\exp \left(2\frac{A-a}{\lambda}\lambda -\lambda\right) \right) \right]\quad \mbox{(using result (ii.))}\\
    &=\exp(-\lambda) \left[\exp\left( \frac{(A-a)^2}{\lambda} \right) -\exp(2A - 2a -\lambda)\right],\\
    \text{and } D & = \mathbb{E}_\chi \mathrm{Var}_{\widehat{A}_P\mid \chi} \left[  \prod_{h=1}^{\chi} \left( \frac{\widehat{A}_P^{\,(h)}-a}{\lambda}\right)\right].
\end{align*}

To derive the term for $D$, we first compute the conditional variance as 

\begin{align*}
    \mathrm{Var}_{\widehat{A}_P\mid \chi} \left[  \prod_{h=1}^{\chi} \left( \frac{\widehat{A}_P^{\,(h)}-a}{\lambda}\right)\right] &= \prod_{h=1}^\chi \left[  \mathrm{Var}_{\widehat{A}_P\mid \chi}\left(\frac{\widehat{A}_P^{\,(h)}-a}{\lambda}\right)+ \left(\mathbb{E}_{\widehat{A}_P\mid \chi}\left(\frac{\widehat{A}_P^{\,(h)}-a}{\lambda}\right)\right)^2 \right] - \prod_{h=1}^\chi \left(\mathbb{E}_{\widehat{A}_P\mid \chi}\left(\frac{\widehat{A}_P^{\,(h)}-a}{\lambda}\right)\right)^2 \\
    &= \left[ \frac{\sigma^2_{\widehat{A}}}{\lambda^2} + \left(\frac{A-a}{\lambda}\right)^2 \right]^\chi-\left[ \left(\frac{A-a}{\lambda}\right)^2 \right]^\chi.
    \end{align*}

Substituting this term into $D$, we have    
    \begin{align*}
    D &= \mathbb{E}_\chi \left[ \left( \frac{\sigma^2_{\widehat{A}}}{\lambda^2} + \left(\frac{A-a}{\lambda}\right)^2 \right)^\chi \right] - \mathbb{E}_\chi\left[ \left(\frac{A-a}{\lambda}\right)^{2 \chi} \right]\\
    &=\exp \left[ \left(\frac{\sigma^2_{\widehat{A}}}{\lambda^2} + \left(\frac{A-a}{\lambda} \right)^2 -1 \right) \lambda \right] -\exp \left[ \left( \left(  \frac{A-a}{\lambda} \right)^2 -1 \right) \lambda \right]\\
    & =\exp \left[\frac{(A-a)^2}{\lambda} -\lambda \right] \left[\exp \left(\frac{\sigma^2_{\widehat{A}}}{\lambda} \right) - 1\right].
\end{align*}
We then have, 
\begin{align*}
    \mathrm{Var}(\widehat{\exp(A_P)}) & =\exp(2a + 2\lambda) \times \\ & \left\{\exp(-\lambda) \left[\exp\left( \frac{(A-a)^2}{\lambda} \right) -\exp(2A - 2a -\lambda)\right] 
    - \exp \left(\frac{(A-a)^2}{\lambda} -\lambda \right) \left[\exp \left(\frac{\sigma^2_{\widehat{A}}}{\lambda} \right) - 1\right] \right\} \\
    &=\exp\left(  \frac{(A-a)^2+\sigma^2_{\widehat{A}}}{\lambda}+\lambda + 2a  \right) -\exp(2A).
\end{align*}
\end{proof}

\begin{proof}{Property 3}

To get the optimal value for $a$, take the first derivative of $\log\left(\mathrm{Var}(\widehat{\exp(A_P)})\right)$ with respect to $a$ and set to zero.
\begin{align*}
    \frac{\partial }{\partial a} \log \left( \mathrm{Var}(\widehat{\exp(A_P)} \right)  =  -2\frac{A-a}{\lambda} +2  = 0,
\end{align*}
which gives $a = A -\lambda$. The second derivative is $\dfrac{2}{\lambda} >0$, confirming the minimum is achieved.
\end{proof}

\section{The separation-of-variables (SOV) estimator}

\label{app: sov}

The general idea behind the SOV estimator is the representation
\begin{align}\label{eq: MV_sampler}
    \int_{B_1} g_{X}(z;\theta) dz =  \int_{B_1-\mu} \phi(z; {0}_d,\Sigma) dz 
    &=(2\pi)^{-d/2}  \int^{u_1'}_{l_1'} \exp\left(-\frac{y_1^2}{2}\right) dy_1 \dots \int^{u_d'}_{l_d'} \exp\left(-\frac{y_d^2}{2}\right) dy_d,
\end{align}
where 
$B_1-\mu=\{x-\mu:x\in B_1\}$ is the bin $B_1$ with a location shift of $\mu$, 
${0}_d$ is a vector of zeros,
$y=( y_1,\dots, y_d)^\top$ is the $d$-vector 
$y = L^{-1}z$, and $\Sigma = LL^\top$ where $L$ is a lower triangular matrix.
 
Denote the marginal lower and upper bounds of the bin $B_1-\mu$ in margin $k$ by $(l_i,u_i)$, $i=1,\ldots,d$, and define $(l'_i, u'_i)$ as
\begin{align*}
    l_1' & = l_1; \quad  u_1' = u_1; \\
    l_i' &= l_{i} - \sum_{j=1}^{i-1} L_{ij} y_j / L_{jj};  \quad  u_i' = u_{i} - \sum_{j=1}^{i-1} L_{ij} y_j / L_{jj},  \quad i = 2,\dots,d,
\end{align*}
where $l_i' \leq y_i \leq u_i'$ for $i = 1,\dots, d$, and $L_{ij}$ denotes the $(i,j)$th element of $L$. The SOV estimator first evaluates the integral  
$$\int^{u_1'}_{l_1'} \exp\left(-\frac{y_1^2}{2}\right) dy_1,$$
and then samples $y_1$ proportional to this density between $l_1'$ and $u_1'$, which is equivalent to sampling from a truncated normal distribution. The region $l_2',u_2'$ is then determined based on the sampled $y_1$, and the next integral $\int^{u_2'}_{l_2'} \exp\left(-\frac{y_2^2}{2}\right) dy_2$  evaluated. 
This process is repeated for all univariate integrals in \eqref{eq: MV_sampler}.
The difference between the MET and the SOV estimators is that SOV uses \eqref{eq: MV_sampler} directly to evaluate the integral, whereas MET uses a tilting parameter to shift the $(l'_i,u'_i)$ as well.  See \cite{botev2016normal} for further details.

\section{Signed block PMMH algorithm}

\label{app: alg2}
Algorithm \ref{algo: PM-MCMC_SDA} shows the signed pseudo-marginal algorithm \citep{lyne2015russian} for symbolic data problems.

\begin{algorithm}[ht!]
\caption{The signed block PMMH algorithm} \label{algo: PM-MCMC_SDA}
\begin{algorithmic}[1]
\State 
\textbf{Input}:

$s$: symbol information.

$u$:  random numbers between 0 and 1, grouped into $U$ blocks.

$\theta_0$: starting value for parameters. 

$iter$: total number of iterations.

\State \textbf{Output}: An unbiased estimator for $\psi(\theta)$ from the target distribution.

\For{$i = 1 \to \textit{iter} $}
\State Generate $u'$ given $u$ by updating one block out of $U$ blocks.
\State Generate $\theta'$ given $\theta_{i-1}$ from $q(\theta'|\theta_{i-1})$.
\State Calculate the acceptance ratio:
\begin{equation}\label{eq: ar_SDA}
\text{acceptance ratio } \alpha = \min \left\{  1, \frac{|\widehat{L}_{B_1}(s;\theta',u')|\pi(\theta') }{|\widehat{L}_{B_1}(s;\theta_{i-1},u)| \pi(\theta_{i-1})}  \times \frac{q(\theta_{i-1}|\theta')}{q(\theta'|\theta_{i-1})}\right\}.
\end{equation} 
\State Generate $a$ from Uniform(0,1).
\If{$\alpha>a$}
\State Accept $\theta_i \gets \theta' $.
\State Update $u \gets u' $.
\Else
\State Maintain $\theta_i \gets \theta $.
\State No update for $u$.
\EndIf
\State $\text{sign}(\theta_i|s) \gets \text{sign}(\widehat{L}_{B_1}(s;\theta_i,u))$. \Comment{$\text{sign} (x) = 1$ if $x > 0$; $\text{sign}(x) = -1$ if $ x < 0$.}
\EndFor
\State $\hat{\psi}(\theta) \gets \dfrac{\sum_{i=1}^{iter} \psi(\theta_i) \text{sign} (\theta_i |s)}{\sum_{i=1}^{iter} \text{sign} (\theta_i|s)}$.
\end{algorithmic}
\end{algorithm}
\end{document}